# Principal component analysis as a tool to extract Sq variation from the geomagnetic field observations: conditions of applicability

## Anna Morozova*,1, Rania Rebbah2

1 University of Coimbra, Instituto de Astrofísica e Ciências do Espaço (IA-U.Coimbra), Coimbra, Portugal,
2 University of Coimbra, CITEUC, Department of Physics, Coimbra, Portugal

## Corresponding author's email address
anna.morozova@uc.pt, annamorozovauc@gmail.com

## Abstract
We analyzed the applicability of the principal component analysis (PCA) as a tool to extract the Sq variation of the geomagnetic field (GMF) taking into account

- different geomagnetic field components,
- data measured at different levels of the solar and geomagnetic activity,
- data from different months.

GMF variations obtained with PCA were "classified" as $Sq_{PCA}$ using reference series:

- obtained from the observational data ($Sq_{IQD}$),
- simulated by ionospheric field models.

The results for the X and Y and Z components are essentially different:

- The Sq variation is always filtered to the first PCA mode for the Y and Z components allowing automatic extraction of Sq with PCA.
- For the X component, the automatic extraction of the Sq is not possible, and a complimentary analysis, like a comparison to a reference series, is always needed.

While our results show that both the data-based and model-based reference series can be used, the DIFI3 model performs better as a reference series for GMF at middle latitudes.

We also recommend to estimate the similarity of the series with a metric that account for possible local stretching/compressing of the compared series, for example, the dynamic time warping (DTW) distance.

## Specifications table

| Subject area | Earth and Planetary Sciences |
|---|---|
| **More specific subject area** | Geomagnetic field variations |
| **Name of your method** | Extraction of the solar quiet (Sq) variation using the principal component analysis (PCA) |
| **Name and reference of original method** | Xu, W.Y. and Kamide, Y. (2004): Decomposition of daily geomagnetic variations by using method of natural orthogonal component. Journal of Geophysical Research: Space Physics, 109(A5).doi: 10.1029/2003JA010216.<br>Chen, G.X., Xu, W.Y., Du, A.M., Wu, Y.Y., Chen, B. and Liu, X.C. (2007): Statistical characteristics of the day-to-day variability in the geomagnetic Sq field. Journal of Geophysical Research: Space Physics, 112(A6), doi:10.1029/2006JA012059.<br>De Michelis, P., Tozzi, R. and Meloni, A. (2009) On the terms of geomagnetic daily variation in Antarctica. Ann. Geophys, 27, pp.2483-2490. |

| | De Michelis, P., Tozzi, R. and Consolini, G. (2010) Principal components' features of mid-latitude geomagnetic daily variation. Ann. Geophys, 28, pp.2213-2226, doi:10.5194/angeo-28-2213-2010. |
|---|---|

**Method details**

**1. Main method to extract Sq variation of the geomagnetic field**

There are three main types of geomagnetic field variations on the time scale from hours to several days: regular variations during a calendar or solar day (so-called "daily" or "solar" variations known as S-type variations), regular variations during a lunar month (L-variation) and irregular variations often associated with storms and substorms and called "disturbances" (Dst-variations), see Chapman and Bartels (1940). The S-type variations are divided into two main classes: the "daily (solar) quiet" variation, Sq, which is observed most clearly during the geomagnetically quiet days, and the "daily (solar) disturbed" variation, SD, (Chapman and Bartels, 1940; Yamazaki and Maute, 2017). More details on the origin and the character of the Sq variation of the geomagnetic field can be found in the Supplementary Material (SM1).

The standard method to obtain Sq from the ground observations of the geomagnetic field consists of the selection of days with the lowest level of geomagnetic field perturbations (so-called "quiet days"), typically, five days per calendar month, and averaging of the daily geomagnetic field variations for a certain component over selected days. These days can be defined using the data of an individual observatory (local quiet days) or using the data from the same set of observatories that are used to calculate the Kp index (international quiet days – IQD), see Chapman and Bartels (1940). Hereafter, the Sq variation obtained using IQD is named "$Sq_{IQD}$".

In this work, we used IQDs routinely provided by the GFZ German Research Centre for Geosciences at the Helmholtz Centre in Potsdam, Germany and available at https://www.gfz-potsdam.de/en/kp-index/ or ftp://ftp.gfz-potsdam.de/pub/home/obs/kp-ap/quietdst/.

The $Sq_{IQD}$ variation for a certain month is calculated as the mean daily variation for five IQDs of a month. Before the averaging, a baseline was removed from the raw daily series of the X, Y and Z components. In this work, the baseline was defined as a mean calculated for the night hours: 00:30 UTC, 01:30 UTC, 02:30 UTC, 03:30 UTC and 23:30 UTC of each analyzed day (for Coimbra UTC = LT). Thus, the $Sq_{IQD}$ variation values for the night hours are close to zero, and there is no significant difference between the night values of Sq at the beginning and the end of a day.

**2. PCA-based methods to extract Sq variation of the geomagnetic field**

Another way to extract regular variations as Sq is to apply a decomposition method to the geomagnetic field data: e.g., the wavelet analysis (Maslova et al., 2010), the empirical mode decomposition (Piersanti et al., 2017) or the principal component analysis, PCA (Xu and Kamide, 2004; Chen et al., 2007; De Michelis et al., 2009, 2010). Also, the shape and position of the vortex can be deduced from the observational data using the spherical harmonic analysis by calculating the equivalent electric field (Takeda, 1982; Haines and Torta, 1994) or it can be reconstructed as equivalent electric current vectors (horizontal component) from the observed horizontal geomagnetic field vector (Stening et al., 2005; Stening, 2008).

First attempts to use PCA (sometimes known as a method of the natural orthogonal component, NOC) to extract regular variations of the geomagnetic field were made in the 1970s-1990s (Golovkov et al., 1978, 1989; Rangarajan and Murty, 1980; Golovkov and Zvereva 1998, 2000) but were not actively supported by the geomagnetic community (Menvielle, 1981). Golovkov et al. (1978, 1989) and Golovkov and Zvereva (1998, 2000) showed that for the H component of the geomagnetic field and for the geomagnetically quiet time intervals, the Sq variation can be associated with the first (or first and third) principal components (PC) and the second PC can be identified as SD variation. For the geomagnetically active time intervals the first PC was identified as SD, and the second and third PCs were identified as Sq. Dependence of the order of a PC that can be identified as Sq or SD on the latitude was also shown. Both the existence of the daily variability of the Sq field and the need for studying it was also emphasized in the early works.

Later, Xu and Kamide (2004) and Chen et al. (2007) revived the interest of the geomagnetic community in PCA as a useful tool that allows not only to extract regular variations of the geomagnetic field, as Sq and SD, but also to analyze seasonal and geographic variations of the phase and amplitude of the Sq and SD fields and the dependence of their intensity on the level of the solar and geomagnetic activity. Works of Wu et al. (2007), De Michelis et al., (2009, 2010), Bhardwaj et al. (2015, 2016) and others (see also review by Yamazaki and Maute, 2017), generally confirmed the applicability of PCA to the extraction of the regular geomagnetic field variations observed at different latitudes, and for the time intervals of different length and corresponding to different geomagnetic activity levels. However, the results obtained for different regions/time intervals were somewhat different.

In particular, it was found that for the H (X) component of the geomagnetic field for the Asian sector (Xu and Kamide, 2004; Chen et al., 2007; Wu et al., 2007; Bhardwaj et al., 2015, 2016) the Sq variation is filtered to the first PC and the SD variation is filtered to the second PC. On the contrary, for the European sector (De Michelis et al., 2010) PC1 is associated with SD and PC2 is associated with Sq. This difference can be explained both by the different geographic positions of the stations whose data were used for PCA and by the different studied time intervals. Also, for the Y (D) and Z components of the geomagnetic field for the European sector PC1 was identified as Sq and PC2 as SD.

To our knowledge, no systematic study of the applicability of PCA as a tool to extract Sq-type variations was performed yet and no possible explanation for the differences mentioned above was proposed. In this work, we present the results of such a study: we test PCA on different components of the geomagnetic field (X, Y and Z), on the data obtained in different months and under different levels of solar and geomagnetic activity. We also tested different lengths of the input data sets.

We use the geomagnetic field data obtained at a European mid-latitudinal geomagnetic observatory – Coimbra Magnetic Observatory (COI) in Portugal. The peculiarity of COI, and this can be also true for the L'Aquila observatory (De Michelis et al., 2010), is that it is located near the mean latitude of the focus of the Sq ionospheric current vortex. Thus, the shape of the Sq variations for the X component at COI can vary not only due to the intensity of the vortex but also due to the position of its focus: for some days COI is located to the north of the focus, for other days it is located to the south of the focus, and there are days when COI is located very near the focus latitude. These changes of the COI relative position result in different shapes of the Sq X variation (see SM1). Finally, contrary to all previous studies, we analyzed the data not on the annual or decadal time scale but on the monthly time scale as described below and in Morozova et al. (2021a, 2021b).

**1. Description of the proposed PCA-based method to extract Sq variation of the geomagnetic field**

Principal component analysis (PCA) allows the extraction of main modes of variability of an analyzed series – principal components or PCs. The full descriptions of this widely used mathematical method can be found in, e.g., Björnsson and Venegas (1997), Hannachi et al. (2007), Shlens (2009). PCs are orthogonal and conventionally non-dimensional. The amplitudes of a PC for each of the analyzed days are given by the corresponding empirical orthogonal function (EOF). The combination of a PC and the corresponding EOF is called a "mode". The "significance" of each of the extracted modes is estimated from the corresponding eigenvalues as variance fraction (VF). VF can be between 0 and 1 and multiplied by 100% shows the per cent of the total variability of the analyzed series related to a particular mode.

The PCA input matrices were constructed as follows. For the individual months and years, the input matrices have 24 rows (24 hourly values per day) and from 28 to 31 columns (a column for a day) depending on the analyzed month. All February matrices have the size 24 x 28. For the individual months but for the "all years" series the input matrices have sizes 24 x 308, 24 x 330 or 24 x 341, depending on the analyzed month. The singular value decomposition (SVD) approach was used to solve the matrix equations.

In this configuration of the input matrices, the principal components (PCs) correspond to daily variations of different types that can be matched up with Sq variation calculated using the standard approach. The amplitudes of PCs for an individual day are given by corresponding EOFs.

Only three first PCs were selected for further analysis. Overall, the first three PCA modes explain together up to 67-94% of the COI X variability, and up to 83-98% of the COI Y and COI Z series variability depending on a month and a year. Table 1 shows VFs associated with the first three PCA modes of the variations of the X, Y and Z components.

During further analyses, PCs were compared to reference series and those PCs that can be classified as Sq were denoted as $Sq_{PCA}$.

Below we provide a systematic analysis of the PCA's performance on mid-latitudinal geomagnetic data for different geomagnetic field components, different seasons and under different levels of the solar and geomagnetic activity. We also tested if only one PC is always sufficient to represent an Sq-type variation or a combination of two PCs should be considered as well.

**Table 1. PCA variance fraction (in %) of the geomagnetic field X, Y and Z components.** The minimum, maximum and mean values of the variance fraction associated with the first three principal components (PC1-PC3) and the cumulative variance fraction (Σ) for the first three PCs. Bold marks PCs that are essential for Sq extraction.

|  | X component | | | Y component | | | Z component | | |
| --- | --- | --- | --- | --- | --- | --- | --- | --- | --- |
|  | min | mean | max | min | mean | max | min | mean | max |
| PC1 | **28.8** | **49.5** | **78.2** | **58.1** | **82.7** | **94.0** | **62.1** | **85.0** | **94.9** |
| PC2 | **9.5** | **21.1** | **36.9** | 1.7 | 6.5 | 22.0 | 1.9 | 6.2 | 17.9 |
| PC3 | **4.2** | **10.9** | **20.7** | 1.1 | 3.5 | 8.5 | 0.8 | 3.1 | 10.9 |
| Σ(PC1 to PC3) | 67.2 | 81.6 | 93.8 | 82.5 | 92.7 | 98.1 | 83.6 | 94.3 | 98.0 |

## 4. Data and methods used for validation

### 4.1 Data

*Geomagnetic field data.* Geomagnetic measurements at the Coimbra Magnetic Observatory in Portugal (40º 13' N, 8º 25.3' W, 99 m a.s.l., IAGA code COI) have been started in 1866 (Morozova et al., 2014, 2021c). The last changes of the instruments took place at COI in 2006: new sets of the absolute instruments were installed providing good quality measurements of geomagnetic field components with 1 hour time resolution (Morozova et al., 2021c). Since that time to the present, there were no changes in the instruments or station location, and the data obtained between 2007 and the present time can be considered homogeneous (Morozova et al., 2021c). A detailed description of the COI instruments and metadata for the series of the geomagnetic field components can be found in Morozova et al. (2014, 2021a, 2021c). The 1h data for all geomagnetic components can be downloaded from the World Data Centre for Geomagnetism using the Geomagnetism Data Portal at http://www.wdc.bgs.ac.uk/dataportal/ (station name: "Coimbra", IAGA code: "COI"). These data were used to obtain both the $Sq_{IQD}$ variation and the main PCA modes of the geomagnetic field variations.

The dataset consists of 1h data on the variations of the X (northern), Y (eastern) and Z (vertical) components of the geomagnetic field measured at COI during 11 years from January 1, 2007, to December 31, 2017. This time interval covers (approximately) one solar cycle. The data for different components were tested separately. The data were used on the time scale of one calendar month. The $Sq_{IQD}$ variation and the PCA modes were calculated for each month both for the individual years, i.e., using only the data for January 2007, for January 2008, etc., separately, and for each month but all years together, i.e., using the data for January 2007 and January 2008, etc. together, hereafter "all years" series. As a result, for each of the three analyzed components, there were obtained 11*12 = 132 series for individual months and years, and 12 "all years" series. This dataset is described in detail in Morozova et al. (2021a) and is available in Morozova et al. (2021b). Standard errors (SE) for the $Sq_{IQD}$ values were calculated for each month relative to the $Sq_{IQD}$ "all years" series.

*Geomagnetic field models.* As reference series (see below in Sec. 6) for the ionospheric field of the X omponent the ionospheric fields generated by two geomagnetic field models, CM5 and DIFI3, were used. The CM5 and DIFI3 reference series were generated for the calendar day 15 of each of 12 months, from January to December. Since for both models the ionospheric field outputs for different years have the same shape but change only in amplitude, the CM5 and DIFI3 reference series ($Sq_{CM5}$ and $Sq_{DIFI3}$, respectively) were used in arbitrary units (a.u.). Detailed descriptions of the models can be found in Sabaka et al. (2002, 2020), and Chulliat et al. (2013, 2016) and Thébault et al. (2016), respectively, and a short summary can be found in the Supplementary Material (SM2).

*Solar and geomagnetic indices.* To estimate the decadal and seasonal variations of the level of the solar and geomagnetic activities we used the following indices. The solar activity was represented by the daily means of the sunspot number series (R) and series of the F10.7 index reflecting variations of the solar UV flux. To see variations of the geomagnetic activity level we used daily means of the Dst, Kp and ap, and AE

geomagnetic indices. All the indices were obtained from the OMNI database at
https://omniweb.gsfc.nasa.gov/form/dx1.html. The daily mean values of these indices were used to
calculate both the monthly means and the IQD means (means calculated using only 5 IQD of a month) for
each of the studied months. Corresponding plots can be found in the Supplementary Material (SM3, Figs.
S3.1-S3.4).

## 4.2. Methods used to classify PCs

The daily variations obtained by PCA (PC1, PC2 and PC3) were compared to the $Sq_{IQD}$, $Sq_{CM5}$ and $Sq_{DIFI3}$ variations and classified, when possible, as $Sq_{PCA}$ using two classification metrics: (1) the absolute value of the Pearson correlation coefficient (r), and (2) a metric called the dynamic time warping distance (dtw). Short descriptions of these metrics are given below.

We tested two approaches to the PCs' classification: allowing the combined classification (either one or a sum of two PCs can be classified as $Sq_{PCA}$) and not allowing the combined classification, i.e., single classification (only one PC per studied month is classified as $Sq_{PCA}$).

The need for the combined classification can be justified by the possibility of PCA to decompose an Sq-type variation into several modes for months when the solar and geomagnetic activities were very low. In such a case an Sq-type variation can be decomposed by PCA into several modes that contain different fine features of Sq.

The sums of PCs were calculated as weighted sums with weights being the monthly mean values of the corresponding EOFs.

For each set of PCs, the classification metrics were calculated between those PCs (or their sums) and the corresponding reference series ($Sq_{IQD}$, $Sq_{CM5}$ or $Sq_{DIFI3}$). Only PC (or a sum of PCs) with metrics that are above (below) a predefined threshold for r (dtw) are used for further classification, and PC (or a sum of PCs) with highest (lowest) values of r (dtw) was classified as $Sq_{PCA}$.

***Correlation analysis and correlation coefficient r.*** Here we used the standard Pearson correlation coefficient. Since in this work we used the SVD method to perform PCA, PCs and EOFs are resolved accurately to a sign. This is because both +1*PC & +1*EOF, and -1*PC & -1*EOF are solutions for an input PCA matrix. There is no general way to solve the sign ambiguity. Keeping this in mind we used the absolute values of the correlation coefficients |r|. The threshold for the classification using the correlation analysis was set as $|r| \geq 0.45$.

The significance of the correlation coefficients was estimated using the Monte Carlo approach with artificial series constructed by the "phase randomization procedure" Ebisuzaki (1997). The obtained statistical significance (p value) considers the probability of a random series to have the same or higher |r| as in the case of a tested pair of the original series.

***Dynamic time warping and the dtw metric.*** When using the correlation coefficient as a measure of similarities between two series one must remember that its value is mostly affected by the similarity of main features existing in the compared series. It may be not sensitive to small-scale features or non-systematic shifts of the local minima or maxima (systematic shifts of the local minima and maxima or a relative shift of a whole series can be accounted for by the lagged correlation analysis). Thus, we would need a metric that is sensitive to irregular deformation of series.

The dynamic time warping (DTW) is a popular metric for comparing time series that is insensitive to local compression and stretches allowing to optimally deforms one of the two input series onto another and calculate a certain measure for the "distance" (dtw) between the studied series (Giorgino, 2009). The smaller the "distance" the higher the similarity between the series, contrary to the correlation coefficient which is higher in the absolute value for the similar series. A description of the DTW algorithm can be found in Giorgino (2009), see also reference herein.

In short, when the similarity of two-time series is studied, one of the series is taken as a "reference" and another is locally stretched or compressed to make it resemble the "reference" as much as possible. The distance (dtw value) between the two series is computed after all stretching/compressing are finished by summing the distances of individual aligned elements. Several DTW algorithms have been proposed in the 1970s in the context of speech recognition (Giorgino, 2009).

The dtw parameter, contrary to r, is not defined on a certain absolute scale. To be able to compare the r and dtw values we had to (1) use standardised (zero mean and unity standard deviation) series to perform the DTW analysis and (2) to compare r and dtw sets obtained for the same pairs of series to see if there is any correspondence between the values of r and dtw. In our tests, it was found that this correspondence can be well fit by Eq. 1

$$dtw = \frac{A(1-r)}{B+(1-r)}, \qquad (1)$$

where A and B are fitting coefficients. An example of a dtw fit on r is shown in the Supplementary Material (SM4, Fig. S4.1) and Table S4.1 presents dtw values for certain values of r when PCs series are compared to different reference series.

The mean dtw threshold that is equivalent to $|r| \geq 0.45$ is $dtw \leq 0.58$ but for individual pairs of PCs vs a reference series, it varies from 0.57 to 0.62. As is shown below in Sec. 6 the DTW analysis allows for a better estimation of the similarity of the studied series than the correlation analysis, and the number of the classified series using this dtw threshold is higher than the number of the classified series using the correlation analysis with the $|r| \geq 0.45$ threshold.

## 5. Performance of PCA as a tool to extract Sq
### 5.1. Performance of PCA for the Y and Z components

Figure 1 shows examples of the first PCs together with $Sq_{IQD}$ for the Y (top panel) and Z (bottom panel) components. The $Sq_{IQD}$ series of the Y and Z components have very stable and specific shapes (see, for example, the width of the $Sq_{IQD}$ ±SE bands in Fig. 1): the $Sq_{IQD}$ Z variation is symmetric around the local noon, while $Sq_{IQD}$ Y is anti-symmetric. These shapes agree well with the shapes of the Sq variations for the Y and Z components expected at a mid-latitudinal geomagnetic station (see SM1). The series for PC1s-PC3s and $Sq_{IQD}$ for all months and all years can be found in Morozova et al. (2021b).

The comparison of PC1-PC3 obtained for the Y and Z components with corresponding $Sq_{IQD}$ using the correlation analysis shows that all the PC1 series for both components can be reliably classified as $Sq_{PCA}$. Figure 2 shows tile plots of the classification of PC1s for Y and Z with numbers showing values of the correlation coefficients (all shown $|r| \geq 0.88$, all p value < 0.01). PC2s of the Y and Z series rarely have a significant correlation with $Sq_{IQD}$ (only 1 case out of 144 for Y and Z, respectively, $|r| = 0.48$-$0.55$, p value > 0.2), and no PC3 has such correlations (corresponding plots can be found in the Supplementary Material (SM5, Figs. S5.1-S5.4). The use of the combined classification does not significantly improve the classification of PCs: the addition of other PCs to PC1 increases the r values insignificantly (please compare Fig. 2 and Figs. S5.1-S5.2 with Figs. S5.3-S5.4 in SM5). Therefore, the combined classification is not needed in the case of the Y and Z components.

Thus, for the Y and Z components for all analyzed months and all 11 years from 2007 to 2017, the PC1 series are defined as $Sq_{PCA}$. This means that Sq is the dominant variation for these components. This also means that for the Y and Z components the probability for Sq variation to be extracted as PC1 is 100%, and, therefore, PCA can be used as a reliable method to extract Sq variations from the Y and Z series when the use of IQD is not possible or not applicable for some reason (e.g., gaps in the analyzed series of the geomagnetic measurements or the overall high geomagnetic activity level of the studied months). Also, for the Y and Z components, the PCA performance does not depend on the season or the level of the solar/geomagnetic activity.

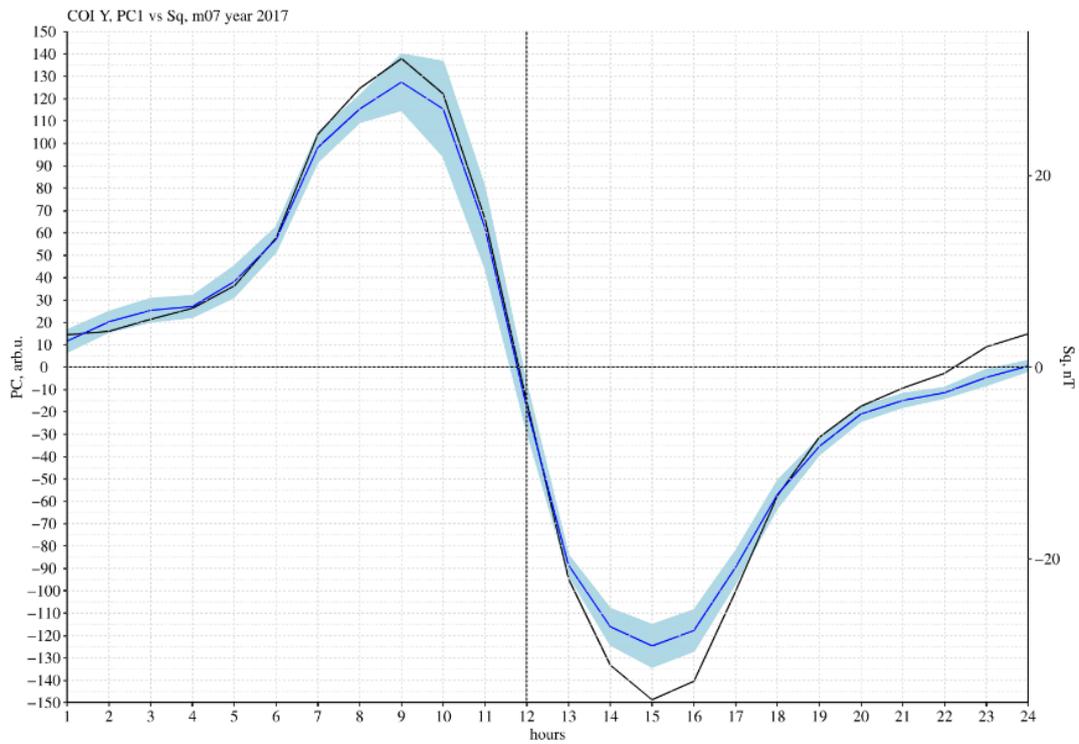

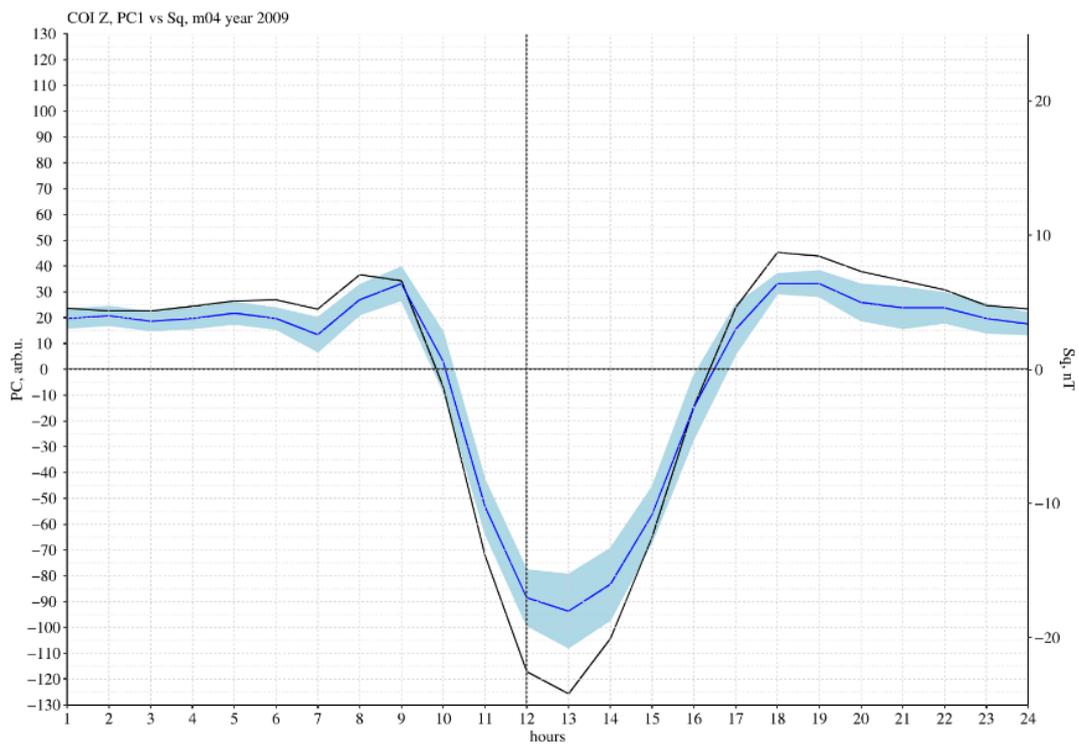

**Figure 1.** Examples of Sq$_{IQD}$ and PC1 daily variations for Y and Z: Sq$_{IQD}$ (blue lines, in nT) and PC1 (black lines, in a.u.) daily variations for the Y (top) and Z (bottom) components. Light blue bands show Sq$_{IQD}$ ± SE values.

## COI Y PC1 classification by r

| months\years | 2007 | 2008 | 2009 | 2010 | 2011 | 2012 | 2013 | 2014 | 2015 | 2016 | 2017 | all |
|---|---|---|---|---|---|---|---|---|---|---|---|---|
| 12 | 0.95 | 0.97 | 0.98 | 0.99 | 0.99 | 0.98 | 0.97 | 0.97 | 0.9 | 0.88 | 0.9 | 0.97 |
| 11 | 0.96 | 0.97 | 0.98 | 0.97 | 0.99 | 0.97 | 0.98 | 0.99 | 0.92 | 0.98 | 0.84 | 0.98 |
| 10 | 0.97 | 0.99 | 0.98 | 0.91 | 0.97 | 0.98 | 0.98 | 0.98 | 0.91 | 0.91 | 0.96 | 0.97 |
| 9 | 0.99 | 0.98 | 0.99 | 1 | 0.98 | 0.94 | 0.96 | 0.99 | 0.97 | 0.97 | 0.97 | 0.99 |
| 8 | 0.98 | 0.99 | 0.99 | 0.99 | 0.99 | 0.97 | 0.98 | 0.99 | 0.99 | 0.98 | 0.98 | 0.99 |
| 7 | 0.99 | 1 | 0.98 | 0.99 | 0.99 | 0.99 | 0.98 | 1 | 0.98 | 0.98 | 1 | 1 |
| 6 | 0.99 | 0.99 | 1 | 0.99 | 0.98 | 0.99 | 0.99 | 0.99 | 0.99 | 0.95 | 0.98 | 0.99 |
| 5 | 0.98 | 0.99 | 0.99 | 0.98 | 0.99 | 0.99 | 0.98 | 0.99 | 0.99 | 0.98 | 0.99 | 0.99 |
| 4 | 0.98 | 0.97 | 0.99 | 0.97 | 0.99 | 0.96 | 0.99 | 0.98 | 0.99 | 0.95 | 0.98 | 0.99 |
| 3 | 0.97 | 0.95 | 0.97 | 0.99 | 0.93 | 0.97 | 0.98 | 1 | 0.96 | 0.95 | 0.95 | 0.98 |
| 2 | 0.94 | 0.91 | 0.88 | 0.92 | 0.92 | 0.98 | 0.97 | 0.92 | 0.97 | 0.95 | 0.9 | 0.96 |
| 1 | 0.88 | 0.92 | 0.95 | 0.95 | 0.98 | 0.97 | 0.96 | 0.96 | 0.93 | 0.93 | 0.92 | 0.97 |

## COI Z PC1 classification by r

| months\years | 2007 | 2008 | 2009 | 2010 | 2011 | 2012 | 2013 | 2014 | 2015 | 2016 | 2017 | all |
|---|---|---|---|---|---|---|---|---|---|---|---|---|
| 12 | 0.94 | 0.97 | 0.97 | 0.93 | 0.99 | 0.99 | 0.96 | 0.95 | 0.81 | 0.87 | 0.89 | 0.96 |
| 11 | 0.97 | 0.98 | 0.97 | 0.97 | 1 | 0.97 | 0.99 | 0.99 | 0.96 | 0.96 | 0.95 | 0.99 |
| 10 | 0.98 | 0.99 | 0.99 | 0.95 | 0.99 | 0.99 | 0.99 | 0.97 | 0.96 | 0.97 | 0.98 | 0.99 |
| 9 | 0.96 | 0.97 | 0.99 | 0.99 | 0.95 | 0.96 | 0.97 | 0.97 | 0.97 | 0.97 | 0.92 | 0.98 |
| 8 | 0.97 | 0.99 | 0.99 | 0.99 | 0.99 | 0.98 | 0.98 | 0.97 | 0.97 | 0.96 | 0.99 | 0.99 |
| 7 | 0.99 | 0.99 | 0.98 | 0.99 | 0.98 | 0.95 | 0.98 | 1 | 0.98 | 0.98 | 0.99 | 0.99 |
| 6 | 1 | 0.99 | 0.99 | 0.99 | 0.99 | 0.99 | 0.95 | 0.99 | 0.98 | 0.98 | 0.97 | 0.99 |
| 5 | 0.98 | 0.99 | 0.99 | 0.97 | 0.99 | 0.99 | 0.98 | 0.99 | 0.99 | 0.98 | 0.99 | 0.99 |
| 4 | 0.98 | 0.98 | 1 | 0.98 | 0.99 | 0.97 | 0.99 | 0.99 | 1 | 0.96 | 0.98 | 0.99 |
| 3 | 0.99 | 0.95 | 0.99 | 0.99 | 0.97 | 0.98 | 0.96 | 1 | 0.97 | 0.96 | 0.92 | 0.98 |
| 2 | 0.97 | 0.94 | 0.91 | 0.98 | 0.97 | 0.97 | 0.97 | 0.95 | 0.99 | 0.97 | 0.98 | 0.98 |
| 1 | 0.9 | 0.91 | 0.84 | 0.94 | 0.97 | 0.96 | 0.92 | 0.96 | 0.93 | 0.92 | 0.94 | 0.96 |

**Figure 2.** Correlation coefficients between the $Sq_{IQD}$ and PC1 series for Y (top) and Z (bottom). Numbers show correlation coefficients for different months (Y-axis) and different years (X-axis). Blue tiles mark the PCs classified as Sq (single classification using r).

Also, using PCA we can estimate a part of the variability of the original Y and Z series associated with the Sq variation. In the case of the validation dataset, as follows from Tab. 1, the mean variance fraction for PC1 for the Y and Z components is ~84%.

It is also possible to detect seasonal variations of VF associated with PC1: in the presented case it is higher during the summer months and lower during winter. These seasonal variations of VF are not driven by the part of the geomagnetic activity, which is described by the Kp, ap or Dst indices: these indices have semi-annual cycles (see Fig. S2.3-S2.4). On the other hand, the AE index describing the geomagnetic activity related to the high-latitudinal magnetosphere and ionosphere has an annual cycle with a maximum in summer (see Figs. S2.3-S2.4). However, to our mind, the main reason for an increase of VF for the first PCA mode during summer is the overall increase of the insolation and the intensification of the Sq current vortex during the summer months (Yamazaki and Maute, 2017).

On the decadal timescale, VF of mode 1 anti-correlates with geomagnetic activity, whereas VFs for mode 2 and mode 3 correlate with geomagnetic activity level (see Tab. 2). This is expected since PC1s for the Y and Z components are associated with Sq, while, consequently, PC2 and PC3 contain variations related to disturbances (e.g., SD and Dst): during years with higher geomagnetic activity the contribution of the disturbance-type variations to the total variability of the Y and Z components increases resulting in higher VF values.

These results agree with previous findings of Golovkov et al. (1978, 1989), Golovkov and Zvereva (1998, 2000) and De Michelis et al. (2010) for different epochs and latitudinal zones.

**Table 2. Correlation between VF and the solar and geomagnetic activity.** The correlation coefficients are calculated between the mean VF associated with a PC for the Y, Z and X components for a certain year, and the corresponding mean values of the solar and geomagnetic indices. Only $|r| \geq 0.3$ are shown, with *p values* in parentheses (only *p values* $\leq 0.2$ are shown). Statistically significant correlation coefficients ($p$ values $\leq 0.05$) are in bold.

|   |   | Geomagnetic indices | | | | Solar indices | |
|---|---|---|---|---|---|---|---|
|   |   | AE | ap | Kp | Dst | R | F10.7 |
| Y component | PC1 | -0.71 (0.15) | -0.78 (0.06) | -0.74 (0.09) |   |   |   |
|   | PC2 | 0.57 | 0.66 (0.13) | 0.61 (0.19) |   |   |   |
|   | PC3 | **0.81 (<0.01)** | **0.81 (0.01)** | **0.75 (0.03)** | -0.44 |   |   |
| Z component | PC1 | -0.50 | -0.56 (0.2) | -0.50 |   | 0.43 | 0.44 |
|   | PC2 | 0.64 (0.09) | **0.69 (0.05)** | 0.63 (0.09) |   |   |   |
|   | PC3 |   |   |   |   | **-0.54 (0.04)** | **-0.59 (0.02)** |
| X component | PC1 |   |   |   |   | 0.46 (0.17) | **0.54 (0.05)** |
|   | PC2 | -0.41 (0.12) | **-0.49 (0.04)** | -0.43 (0.13) | 0.31 |   |   |
|   | PC3 |   |   |   |   |   |   |

## 5.2. Performance of PCA for the X component

Figure 3 shows examples of PC1, PC2 and PC3 together with Sq$_{IQD}$ for the X component. All PC1, PC2 and PC3 series as well as the Sq$_{IQD}$ series can be found at Morozova et al. (2021b). There are two main types of the shape of the Sq X variations obtained from the COI data:

- Shape A: the curves with a minimum (or maximum) near the local noon and secondary (with a lower amplitude) maximum (or minimum, respectively) in the early morning or late afternoon (see Fig. 3a).
- Shape B: the curves with two minima and two maxima of comparable amplitudes (see Fig. 3b-c).

According to Amory-Mazaudier (1994, 2001 and 2009), and Anad et al. (2016), these two types of the Sq X shape can be interpreted, e.g., as caused by an Sq current vortex with a focus located to the south (or to the north, respectively) of the COI location - shape A, or very close to the latitude of COI (40ºN) - shape B.

***Classification of X component PCs by correlation analysis.*** The comparison of PC1-PC3 obtained for the X component with corresponding Sq$_{IQD}$ using the correlation analysis shows that, contrary to the Y and Z components, no PC is always classified as Sq. Figures 4-6 show the classification of PCs for the X component based on the correlation analysis using single and combined classification.

*Single classification (Fig. 4)*: for the individual years' series PC1s and PC2s were classified as Sq at about the same rate (59 and 52 series, respectively, or 40-45% each) while PC3s are classified as Sq about three times less often (18 series or 14%). Only in 3 cases (2%) none of the first three PCs was classified as Sq. On the contrary, for the "all years" series (see Fig. 4, last columns) in 6 cases (50%) PC2s were classified as Sq and in 3 cases each either PC1s or PC3s (25% each) were classified as Sq.

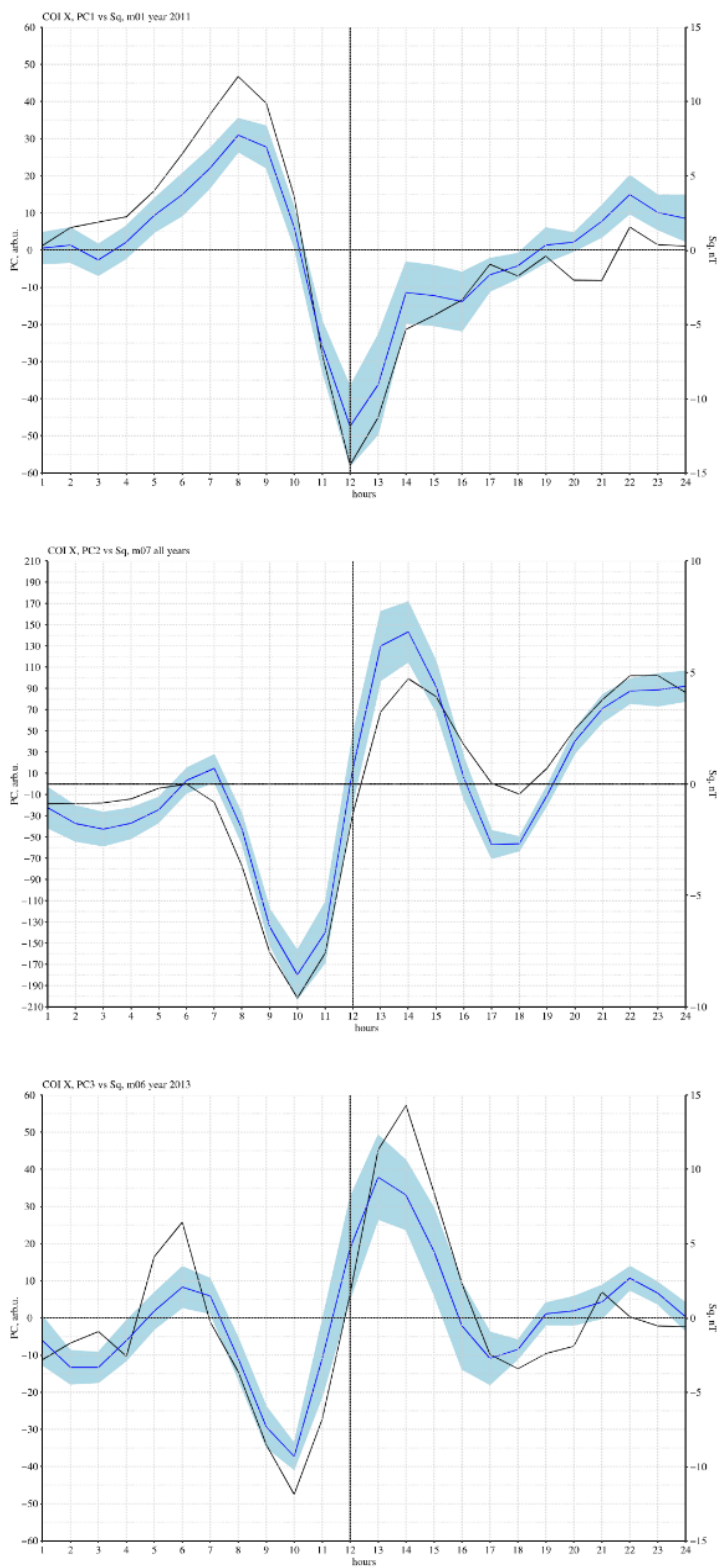

**Figure 3.** Examples of Sq$_{IQD}$ and PC1 daily variations for X. Sq$_{IQD}$ (blue lines, in nT) and PC1 (top), PC2 (middle) and PC3 (bottom) daily variations (black lines, in a.u.) for the X component. Light blue bands show Sq$_{IQD}$ ± SE values.

Thus, for the single classification, the probabilities of PC1 or PC2 to be classified as $Sq_{PCA}$ (or the probabilities of Sq-type variation to be filtered to the 1st or 2nd mode) are approximately equal and about three times higher than the probability of PC3 to be classified as Sq.

*Combined classification (Figs. 5-6):* for the individual years' series PC1s and PC3s were classified as Sq at the same rate (8 and 6 series, respectively, or 5-6%) while PC2s are classified as Sq about two-three times more often (15 series or 12%). The combinations of PCs were classified as Sq in 43 cases for PC1+PC2 (33%), in 25 cases for PC1+PC3 (19%) and 32 cases for PC2+PC3 (24%). Only in 2 cases (~1.5%) none of the first three PCs or their combination was classified as Sq. For the "all years" series (see Figs. 5-6, last columns) in 7 cases (58%) PC2+PC3 were classified as Sq, in 4 cases (33%) PC1+PC3 were classified as Sq, in 1 case (8%) PC2 was classified as Sq.

Thus, for the combined classification the most probable scenarios to extract Sq-type variations are (in the declining order) the combination of PC1+PC2, PC2+PC3 and PC1+PC3.

The results of both kinds of classification for the X component are in general agreement with previous results obtained for the European region (De Michelis et al., 2010): $Sq_{PCA}$ tends to be more frequently associated with PC2 than with other components.

The advantage of the combined classification is that the higher values of the correlation coefficients r were obtained for sums of PCs comparing to r for the individual PCs. In many cases the increase of the r values is small, however in some cases the use of a sum of PCs allows to increase the r value, for example, from 0.6-0.68 to 0.83-0.91 (the cases of June "all years" series, June 2009, July 2017, April 2015, or December "all years" series).

As follows from Tab. 1, the mean variance fractions for PC1, PC2 and PC3 for the X component are ~50%, ~21% and ~11%, respectively. The mean VF varies throughout the year: for PC1 it is higher in winter, and VFs of PC2 and PC3 are higher in summer. On the decadal time scale, see Tab. 2, VF of PC1 (PC2) correlates (anti-correlates) with variations of the geomagnetic activity through the 11-year cycle.

As one can see from Figs. 4-6, there is no clear seasonal or decadal pattern in the classification of PC1-PC3 for the X component as the Sq variation. We compared the number of months per year with PC1, PC2 or PC3 classified as Sq (single classification) with the annual mean values of the solar and geomagnetic activity indices. For the combined classification we made a similar comparison for the number of months per year with PC1+PC2, PC1+PC3 or PC2+PC3 classified as Sq (the number of single PCs classified as Sq using the combined classification is too small for a statistically significant analysis).

The obtained correlation coefficients (Tab. 3) are low and statistically insignificant (all p values > 0.2); however, we may conclude that, in general, the increase of the geomagnetic activity results in a more often classification of PC2 or PC1+PC3 as Sq variation; the increase of the solar activity results in a more often classification of PC1 or PC1+PC3 as Sq. We can interpret this as follows: for geomagnetically quiet epochs the Sq variation is, in most cases, the dominant variation for the X component and has a high probability to be filtered by PCA to the mode 1, while for the geomagnetically disturbed epochs the disturbance-type variations (like SD and Dst) became dominant and will be associated with PC1 while Sq will be rather filtered to the mode 2 or even to the mode 3. Similar behavior was shown by Golovkov et al. (1978, 1989) and Golovkov and Zvereva (1998, 2000) for data obtained at other latitudinal zones and for other decades. On the other hand, the increase of the solar activity results in a more intense flux of the solar UV radiation and, consequently, in higher ionization of the ionosphere, stronger Sq vortex and higher amplitude of the Sq geomagnetic variation. Unfortunately, the found dependence cannot be used to automatically define which PC is classified as Sq.

Thus, for the X component, PCA cannot be used as a simple method to extract Sq variations without further classification of the modes, and a comparison to a reference series is needed to identify PC that represents Sq variation.

## COI X PC1 classification by r

| months \ years | 2007 | 2008 | 2009 | 2010 | 2011 | 2012 | 2013 | 2014 | 2015 | 2016 | 2017 | all |
|---|---|---|---|---|---|---|---|---|---|---|---|---|
| 12 | 0.23 | **0.7** | 0.18 | 0.15 | **0.91** | **0.75** | 0.49 | **0.74** | 0.48 | 0.1 | 0.53 | 0.49 |
| 11 | 0.61 | **0.79** | 0.09 | **0.65** | **0.79** | 0.29 | **0.79** | **0.82** | **0.82** | 0.4 | 0.48 | **0.67** |
| 10 | 0.17 | 0.25 | 0.59 | 0.01 | 0.06 | 0.43 | 0.38 | **0.75** | 0.31 | 0.54 | 0.43 | 0.45 |
| 9 | **0.73** | 0.13 | **0.98** | **0.96** | 0.08 | 0.16 | **0.8** | 0.41 | 0.5 | 0.17 | 0.17 | 0.33 |
| 8 | **0.79** | **0.76** | **0.67** | **0.93** | **0.88** | **0.99** | **0.82** | 0.17 | 0.28 | 0.13 | **0.62** | 0.17 |
| 7 | 0.48 | **0.9** | 0.46 | **0.9** | 0.37 | **0.61** | **0.68** | **0.94** | 0.02 | **0.55** | 0.31 | 0.04 |
| 6 | **0.63** | **0.65** | **0.66** | 0.42 | **0.86** | **0.75** | 0.04 | **0.9** | 0.42 | 0.22 | **0.67** | 0.5 |
| 5 | **0.68** | **0.86** | **0.89** | 0.18 | 0.4 | **0.91** | **0.93** | **0.86** | **0.75** | 0.22 | **0.67** | **0.81** |
| 4 | **0.68** | **0.68** | **0.96** | **0.87** | 0.15 | 0.44 | **0.88** | **0.92** | 0.6 | **0.73** | 0.29 | **0.89** |
| 3 | **0.83** | 0.4 | **0.96** | **0.63** | 0.61 | **0.74** | 0.37 | 0.37 | 0.36 | 0.15 | 0.32 | 0.22 |
| 2 | **0.52** | 0.42 | **0.91** | 0.25 | 0.33 | 0.1 | **0.91** | **0.54** | **0.9** | **0.74** | 0.46 | **0.59** |
| 1 | 0.45 | 0.47 | 0.35 | 0.33 | **0.95** | 0.5 | 0.02 | **0.65** | **0.71** | 0.62 | 0.5 | 0.51 |

## COI X PC2 classification by r

| months \ years | 2007 | 2008 | 2009 | 2010 | 2011 | 2012 | 2013 | 2014 | 2015 | 2016 | 2017 | all |
|---|---|---|---|---|---|---|---|---|---|---|---|---|
| 12 | **0.7** | **0.58** | **0.58** | **0.8** | 0.09 | 0.32 | 0.51 | 0.51 | **0.79** | **0.54** | **0.68** | **0.61** |
| 11 | **0.75** | 0.39 | **0.91** | 0.16 | 0.54 | **0.9** | 0.17 | 0.41 | 0.04 | **0.55** | 0.07 | 0.02 |
| 10 | **0.85** | 0.39 | 0.26 | **0.85** | **0.66** | **0.52** | **0.71** | 0.43 | 0.16 | **0.59** | 0.49 | 0.41 |
| 9 | 0.36 | **0.92** | 0.1 | 0.24 | **0.85** | 0.04 | 0.24 | **0.83** | 0.05 | 0.45 | **0.76** | 0.28 |
| 8 | 0.27 | **0.56** | 0.05 | 0.3 | 0.35 | 0.12 | 0.38 | **0.89** | **0.94** | **0.72** | 0.46 | **0.98** |
| 7 | **0.53** | 0.37 | **0.78** | 0.17 | **0.66** | 0.47 | 0.53 | 0.01 | **0.93** | **0.72** | **0.9** | **0.93** |
| 6 | 0.22 | 0.42 | **0.68** | **0.53** | 0.22 | 0.18 | 0.11 | 0.11 | 0.27 | 0.47 | **0.66** | **0.6** |
| 5 | 0.47 | 0.36 | 0.37 | 0.42 | **0.67** | 0.25 | 0.03 | 0.31 | 0.51 | 0.42 | **0.58** | 0.23 |
| 4 | **0.51** | **0.58** | 0.2 | 0.24 | **0.84** | **0.8** | 0.04 | 0.2 | **0.62** | 0.6 | **0.71** | 0.27 |
| 3 | 0.36 | 0.38 | 0.17 | 0.18 | **0.75** | **0.6** | **0.84** | **0.53** | **0.82** | **0.74** | **0.92** | **0.9** |
| 2 | 0.35 | **0.57** | 0.33 | **0.49** | **0.56** | **0.9** | 0.28 | **0.55** | 0.14 | 0.1 | **0.76** | **0.52** |
| 1 | **0.73** | **0.54** | **0.53** | **0.87** | 0.25 | 0.28 | 0.44 | 0.4 | 0.43 | **0.75** | 0.16 | **0.57** |

## COI X PC3 classification by r

| months \ years | 2007 | 2008 | 2009 | 2010 | 2011 | 2012 | 2013 | 2014 | 2015 | 2016 | 2017 | all |
|---|---|---|---|---|---|---|---|---|---|---|---|---|
| 12 | 0.42 | 0.26 | 0.38 | 0.43 | 0.3 | 0.54 | **0.63** | 0.35 | 0.06 | 0.28 | 0.45 | **0.58** |
| 11 | 0.25 | 0.46 | 0.06 | 0.64 | 0.23 | 0.25 | 0.43 | 0.39 | 0.35 | 0.51 | **0.79** | **0.7** |
| 10 | 0.22 | 0.14 | **0.64** | 0.36 | 0.41 | 0.2 | 0.44 | 0.15 | 0.32 | 0.2 | 0.12 | **0.53** |
| 9 | 0.49 | 0.29 | 0.07 | 0.04 | 0.3 | **0.92** | 0.37 | 0.2 | **0.78** | **0.5** | 0.17 | **0.88** |
| 8 | 0 | 0.22 | **0.72** | 0.06 | 0.27 | 0.04 | 0.38 | 0.3 | 0.07 | 0.56 | 0.48 | 0.07 |
| 7 | 0.16 | 0.09 | 0.17 | 0.14 | 0.6 | 0.6 | 0.45 | 0.02 | 0.14 | 0.34 | 0.02 | 0.26 |
| 6 | **0.69** | 0.28 | 0.17 | **0.65** | 0.27 | 0.25 | **0.9** | 0.32 | 0.03 | **0.76** | 0.03 | **0.59** |
| 5 | 0.47 | 0.29 | 0.18 | **0.72** | 0.45 | 0.03 | 0.28 | 0.34 | 0.29 | **0.54** | 0.42 | 0.47 |
| 4 | 0.02 | 0.35 | 0.03 | 0.29 | 0.38 | 0.32 | 0.12 | 0.13 | 0.05 | 0.03 | 0.32 | 0.3 |
| 3 | 0.02 | **0.54** | 0.06 | 0.53 | 0.18 | 0.2 | 0.34 | 0.02 | 0.14 | 0.54 | 0.13 | 0.35 |
| 2 | 0.07 | 0.39 | 0.17 | 0.15 | **0.66** | 0.3 | 0.1 | 0.42 | 0.38 | 0.13 | 0.24 | 0.42 |
| 1 | 0.22 | 0.51 | 0.01 | 0.14 | 0.15 | **0.51** | **0.8** | 0.47 | 0.48 | 0.11 | **0.59** | 0.51 |

**Figure 4.** Correlation coefficients between the $Sq_{IQD}$ and PCs for X (single classification). Numbers show correlation coefficients between the $Sq_{IQD}$ and PC1 (top), PC2 (middle) and PC3 (bottom) series for the X component for different months (Y-axis) and different years (X-axis). Blue tiles mark PCs classified as Sq (single classification using r).

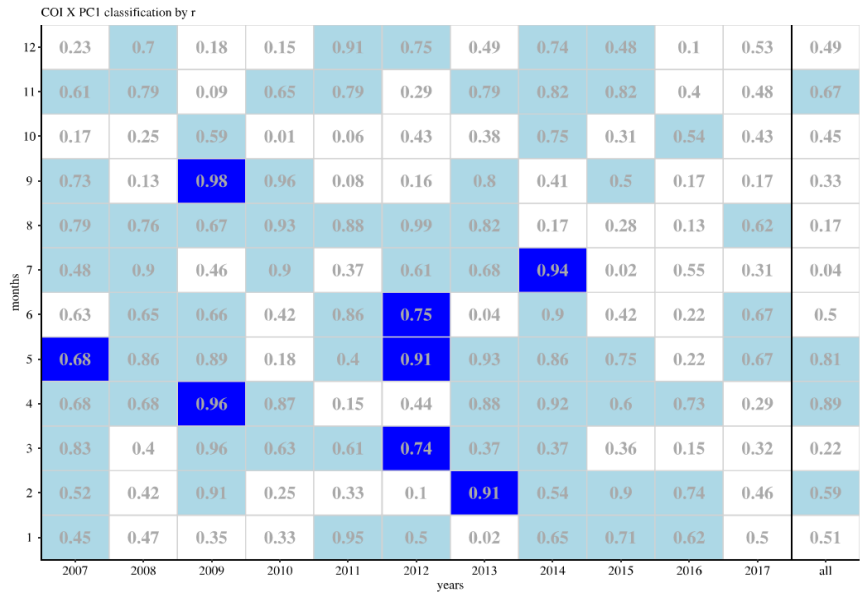
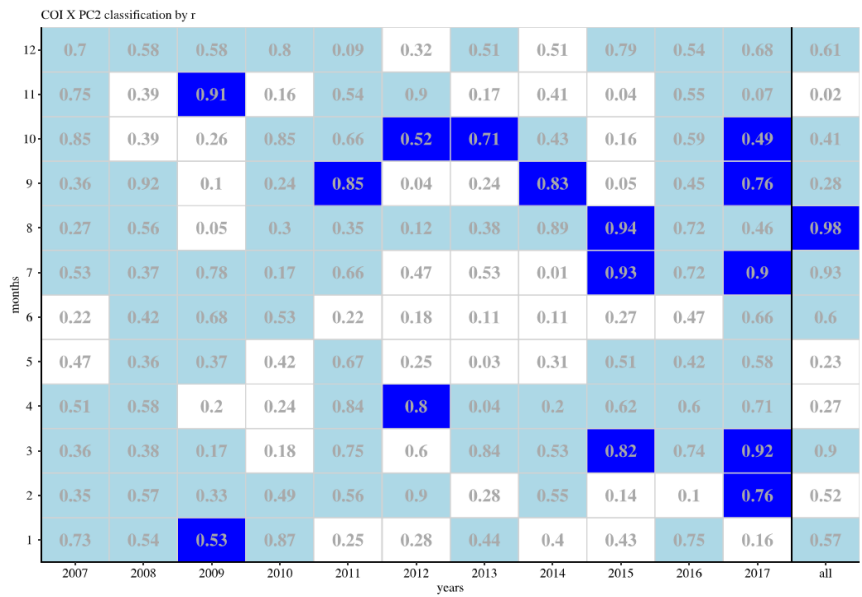
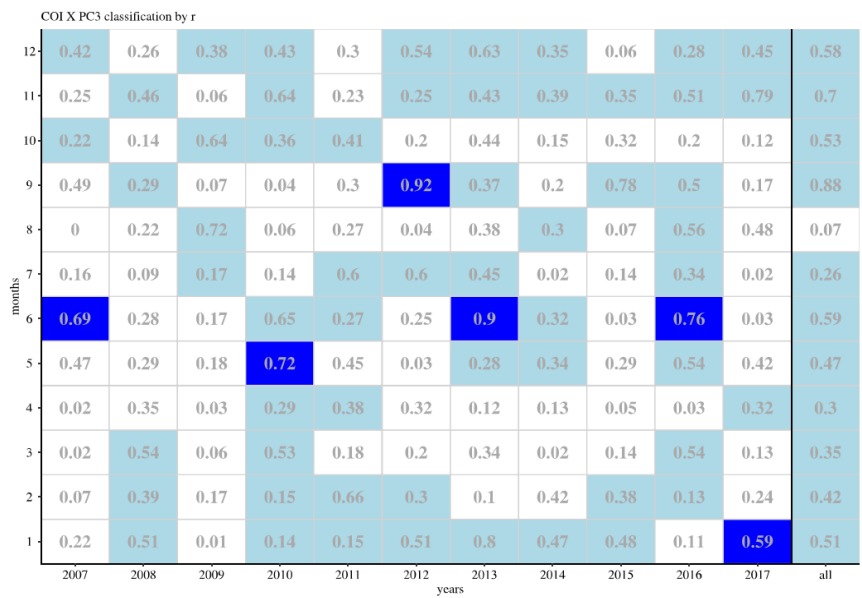

**Figure 5.** Correlation coefficients between the Sq$_{IQD}$ and PCs for X (combined classification). Same as Figure 4 but with the combined classification allowed. Light blue tiles mark PCs classified as Sq in pairs with another PC (see also Figure 6).

**Figure 6**. Correlation coefficients between the Sq$_{IQD}$ and sums of PCs for X (combined classification). Same as Fig. 4 but for sums of PCs: top – PC1+PC2, middle – PC1+PC3, bottom – PC1+PC3.

**Table 3. Number of months with PCs classified as Sq vs mean values of the solar/geomagnetic indices.** Correlation coefficients between the number of months per year with PCs for X classified as Sq and mean annual values of the solar/geomagnetic indices. Only $|r| \geq 0.3$ are shown, with *p values* in parentheses (only *p values* ≤ 0.2 are shown).

|  |  | Geomagnetic indices | | | | Solar indices | |
| --- | --- | --- | --- | --- | --- | --- | --- |
|  |  | AE | ap | Kp | Dst | R | F10.7 |
| Single classification | PC1 | -0.48 | -0.53 (0.18) | -0.47 |  | 0.32 | 0.34 |
|  | PC2 | 0.34 | 0.39 | 0.38 |  |  |  |
| Combined classification | PC1+PC2 |  |  |  | 0.4 | -0.38 | -0.33 |
|  | PC1+PC3 |  |  |  | -0.38 | 0.53 | 0.57 |

## 6. Adaptation of PCA for an automatic extraction of Sq for the X component.

As was shown above, for the X component is it impossible to automatically extract Sq variation using just PCA. A certain reference series is needed to be compared with PCs to classify one of those PCs or a sum of PCs as $Sq_{PCA}$. In this work, we tested two types of reference series: (1) the mean $Sq_{IQD}$ series obtained from geomagnetic field observations for a long time interval and (2) simulations of the ionospheric part of the geomagnetic field using geomagnetic field models.

Another significant problem of the application of PCA to extract Sq-type variation from the series of the X component is that the high rate of the PCs' classification shown above was obtained for a quite low threshold ($|r| \geq 0.45$). As one can see from Figs. 4-6, higher values of the threshold would significantly decrease the number of the identified PCs. On the other hand, the visual analysis of the corresponding PCs and Sq curves shows that in some cases of the low r values the compared series show quite similar variations, and the low values of r are related to local compressions and stretches of one of the series relatively to another. Thus, we need a different metric as a base for the classification. Here we tested the DTW distance as a metric of the similarity of the studied series.

*Mean $Sq_{IQD}$ as a reference series.* $Sq_{IQD}$ series obtained for the same geomagnetic station or observatory seem to be a good choice for a reference series because they automatically incorporate features of the Sq variation associated with a particular location (shape of the daily curve, characteristic seasonal variations etc.). However, as was shown above, the $Sq_{IQD}$ series obtained for an individual month and individual year cannot be used as reliable reference series. Firstly, the automatic usage of PCA implies that a reference series already exists. Secondly, $Sq_{IQD}$ calculated for a particular month and a particular year is strongly affected by the level of geomagnetic activity of those 5 IQDs that were used to calculate it. Thirdly, the position and the shape of the Sq current vortex in the ionosphere depends on the conditions in the upper atmosphere (wind strength, amplitude of waves and tides etc.). Thus, individual features of the vortex during the selected 5 IQDs are preserved in the $Sq_{IQD}$ of the individual months. This can be an essential flaw for an analysis of the data obtained at observatories as COI when the position of the station to the north or the south to the Sq current vortex focus during selected days changes and affects the $Sq_{IQD}$ variation's curve dramatically. On the other hand, averaging the $Sq_{IQD}$ variation series obtained for a certain month but for several years may reduce the effect of individual features caused by the varying geomagnetic and atmospheric conditions. Therefore, we tested the $Sq_{IQD\ allY}$ series, which were calculated for a particular month using data for all years from 2007 to 2017, as one of the reference series for the PCs' classification.

Overall, for most of the studied series (months from January to December, years from 2007 to 2017) there is a strong correlation between $Sq_{IQD}$ and $Sq_{IQD\ allY}$, however, for some months (mostly autumn-winter months with weak Sq current vortex) there is a large variability in the $Sq_{IQD}$ shape resulting in a lower correlation between two types of $Sq_{IQD}$ (individual correlation coefficients can be found in Fig. 7, top). The detailed analysis of the $Sq_{IQD}$ and $Sq_{IQD\ allY}$ series shows that there are (1) cases of low correlation which are caused simply by shifts of maxima/minima position, and (2) cases of (relatively) high correlation that result from the similarity of the general trend but not of individual features of the compared curves. To test if the DTW analysis can perform better in these situations we calculated the dtw values for each of the corresponding pairs of the $Sq_{IQD}$ and $Sq_{IQD\ allY}$ series (Fig. 7, bottom). In general, it seems that the DTW analysis gives a more realistic estimate of the similarity between the $Sq_{IQD}$ and $Sq_{IQD\ allY}$ series. Some examples of the DTW matching can be found in Figs. 8-9: Fig. 8 gives examples of the cases when (relatively) high r values are obtained for series with similar general trends but different local features – corresponding dtw are high which means bad matching between the curves; and Fig. 9 gives examples of the cases when (relatively) low r values are obtained for series with similar features shifted locally – corresponding dtw are low which means good matching between the curves.

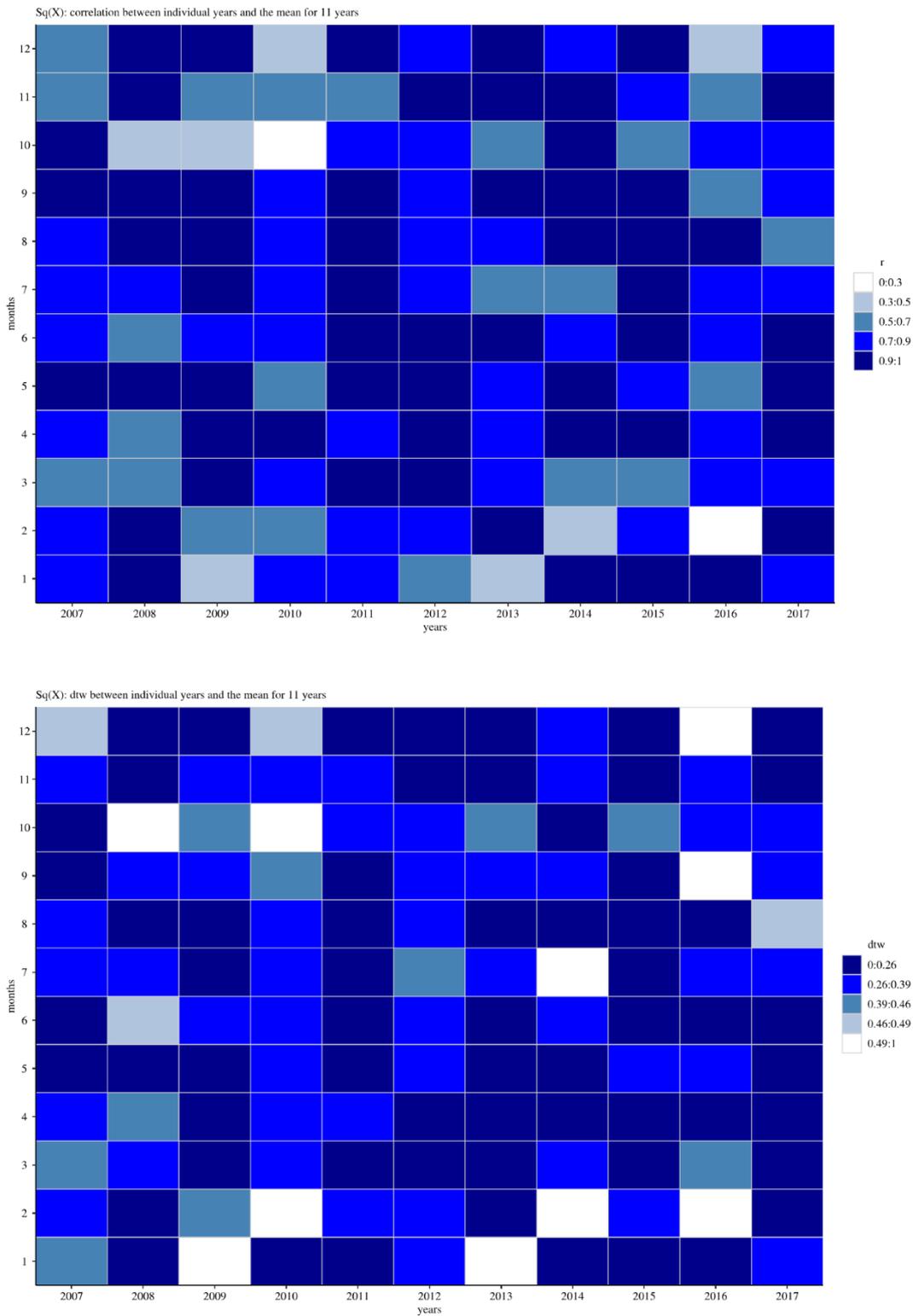

**Figure 7.** Correlation coefficients r (top) and dtw values (bottom) between $Sq_{IQD}$ (calculated for a particular month using data for individual years) and $Sq_{IQD\ allY}$ (calculated for a particular month using data for all years from 2007 to 2017). Colour shows the corresponding ranges of r and dtw.

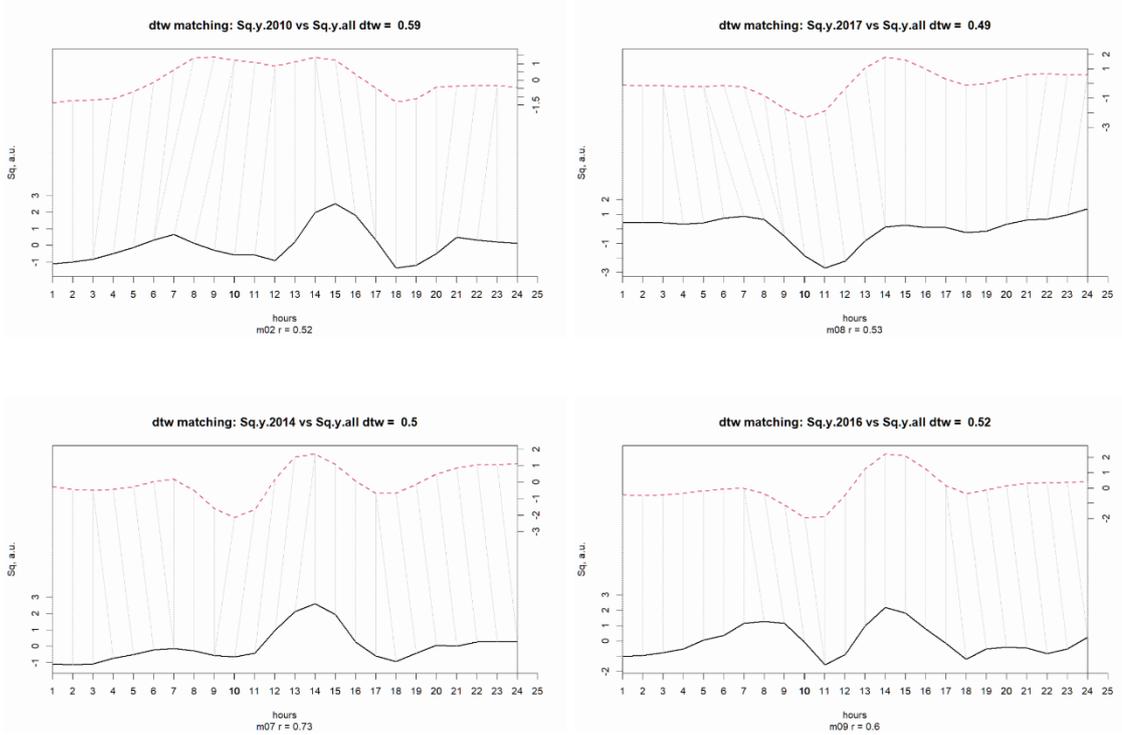

**Figure 8.** Examples of the DTW matching between the Sq$_{IQD}$ (black lines) and Sq$_{IQD\,allY}$ (red lines) when high r does not correspond to low dtw. Corresponding r and dtw values are shown below and above the plots, respectively.

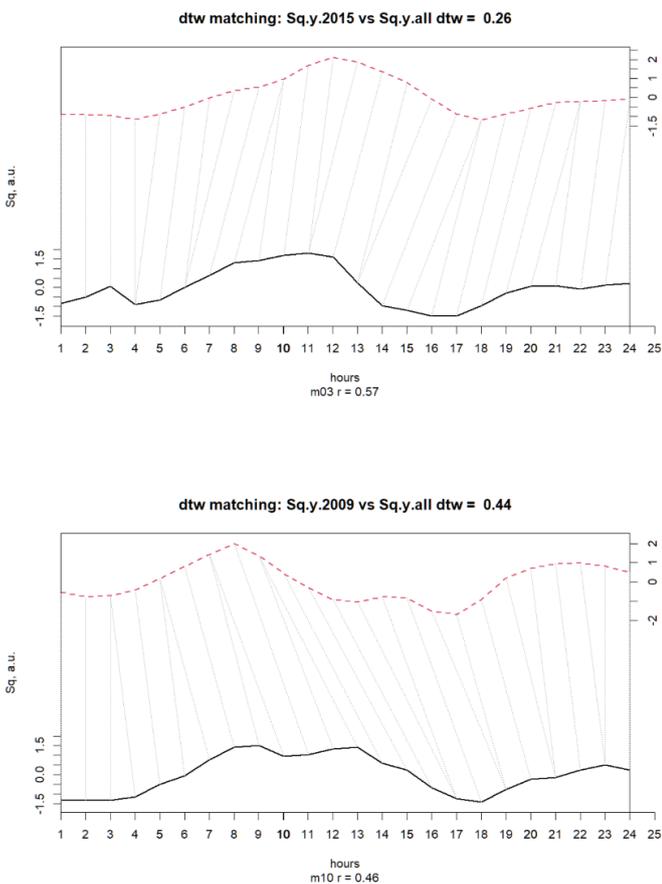

**Figure 9.** Examples of the DTW matching between the Sq$_{IQD}$ (black lines) and Sq$_{IQD\,allY}$ (red lines) when dtw values are much lower than expected for respective r. Corresponding r and dtw values are shown below and above the plots, respectively.

We used the $Sq_{IQD\ allY}$ series as reference series to classify PCs based both on the r and dtw metrics and using the combined classification option. The results (similar to Figs. 5-6) are shown in Figs. 10-11 for r classification and Figs. 12-13 for dtw classification. Columns 1 and 4 of Tab. 4 show how many different PCs or their sums were classified as $Sq_{PCA}$ using r and dtw, respectively.

**Table 4. PCs of X or their sums classified as Sq.** Number (out of 144) of PCs or their sums classified as Sq using different reference series ($Sq_{IQD\ allY}$, CM5 and DIFI3) and different metrics (*r* and *dtw*).

|  | dtw | | | r | | |
|---|---|---|---|---|---|---|
|  | $Sq_{IQD\ allY}$ | CM5 | DIFI3 | $Sq_{IQD\ allY}$ | CM5 | DIFI3 |
| PC1 | 6 | 10 | 11 | 11 | 9 | 7 |
| PC2 | 16 | 17 | 8 | 14 | 15 | 15 |
| PC3 | 12 | 17 | 11 | 7 | 8 | 9 |
| PC1+PC2 | 40 | 52 | 49 | 40 | 68 | 49 |
| PC1+PC3 | 34 | 31 | 31 | 27 | 26 | 28 |
| PC2+PC3 | 35 | 16 | 34 | 43 | 16 | 33 |
| none | **1** | **1** | **0** | 2 | 2 | 3 |

*Ionospheric field models as reference series.* As was mentioned above (Sec. 4.1), we used two models to simulate the ionospheric part of the geomagnetic field: CM5 and DIFI3. These modeled series were used as reference series for the PCs classification using both r and dtw metrics. The classification results (similar to Figs. 5-6 and 10-13) can be found in Figs. 14-15 for r and Figs. 16-17 for dtw for the DIFI3 reference series and in the Supplementary Material (SM6, Figs. S6.1-S6.2 for r and Figs. S6.3-S6.4 for dtw) for the CM5 reference series. Table 4 (columns 2-3 and 5-6) shows the number of different PCs or their sums that were classified as $Sq_{PCA}$ using different models and different metrics.

As one can see from Table 4, dtw allows the classification of more series than r. Most of the series that were identified as $Sq_{PCA}$ are sums of PCs: the sum PC1+PC2 is most often classified as $Sq_{PCA}$ using all studied reference series and both metrics; it is followed by the sums PC2+PC3 and PC1+PC3 which are more or less equally often classified as $Sq_{PCA}$.

While the differences between the performance of the analyzed reference series and metrics are not great, we recommend the DIFI3 model as a reference series and the dtw as a metric to be used to identify PCs that correspond to the Sq-type variations of the X component of the geomagnetic field.

As the final step, we compared the Sq-type variations extracted from the data using PCA and identified using combined classification with DIFI3 as a reference series and dtw as a metric to all reference series ($Sq_{IQD}$, $Sq_{IQD\ allY}$ and $Sq_{DIFI3}$). The mean and median correlation coefficients between $Sq_{PCA}$ and the reference series are $r_{mean} \sim 0.75$ and $r_{median} \sim 0.65$; the individual correlation coefficients can be found in the Supplementary Material (SM7).

Figure 18 shows two examples of the comparisons of $Sq_{PCA}$ and the reference series. These plots also allow comparing $Sq_{PCA}$ variations obtained for a certain month using the data only for a certain year (black lines) and for the "all years" series (grey lines). It seems that for months near equinoxes and solstices (February-March, May-June, August-October, December) it is better to use only data for the studied year to obtain a $Sq_{PCA}$, whereas for other months it is better to use data for this month but for several years of observations (11 in our case), however, this conclusion still needs to be confirmed on longer time series or data from other locations.

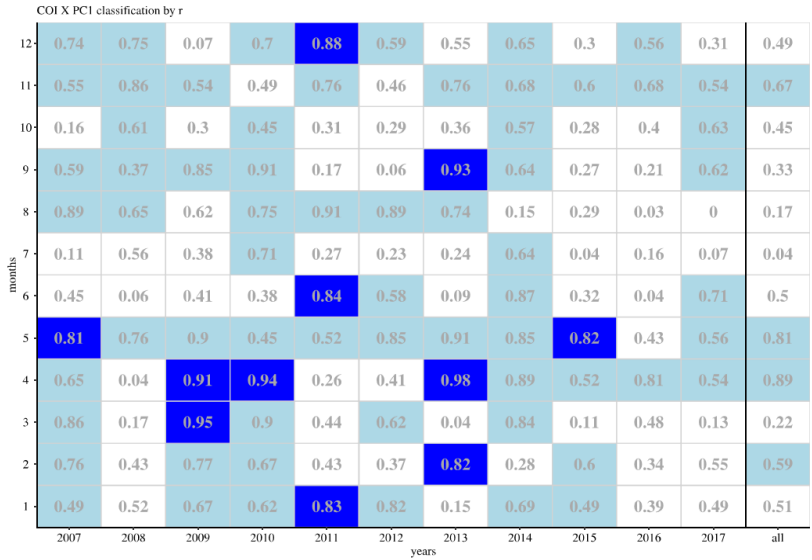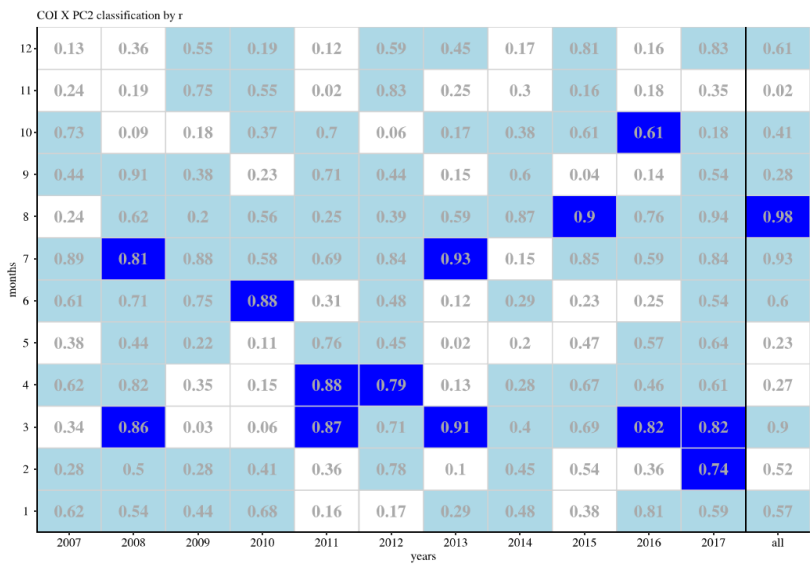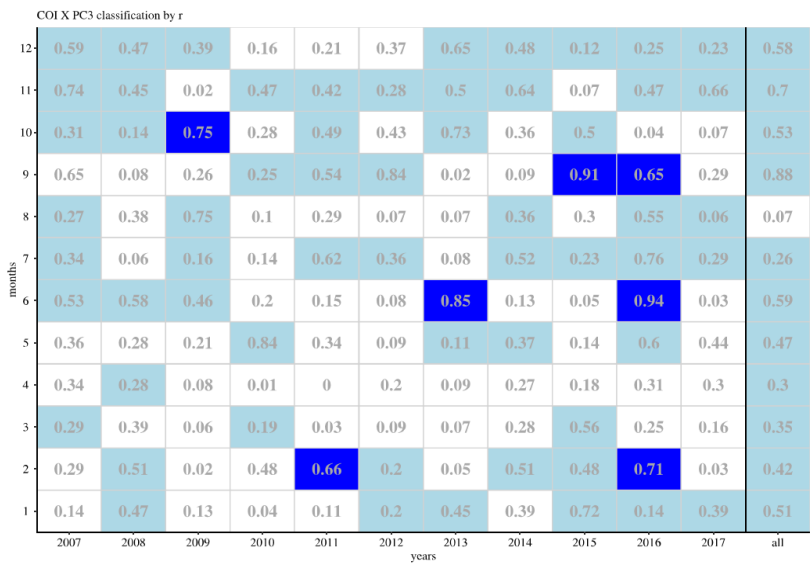

**Figure 10.** Combined classification for the X component with Sq$_{IQD\ allY}$ as a reference series and r is the classification parameter: correlation coefficients (numbers) between the Sq$_{IQD}$ and PC1 (top), PC2 (middle) and PC3 (bottom) series for different months and different years. Blue tiles mark PCs classified as Sq and light blue tiles mark PCs classified as Sq in pairs with another PC (see also Figure 11).

**Figure 11.** Same as Fig. 10 but for a pair of PCs (top – PC1+PC2, middle – PC1+PC3, bottom – PC1+PC3).

**Figure 12.** Combined classification for the X component with Sq$_{IQD\ allY}$ as a reference series and dtw is the classification parameter: dtw values (numbers) between the Sq$_{IQD}$ and PC1 (top), PC2 (middle) and PC3 (bottom) series for different months and different years. Blue tiles mark PCs classified as Sq and light blue tiles mark PCs classified as Sq in pairs with another PC (see also Figure 13).

**Figure 13**. Same as Fig. 12 but for a pair of PCs (top – PC1+PC2, middle – PC1+PC3, bottom – PC1+PC3).

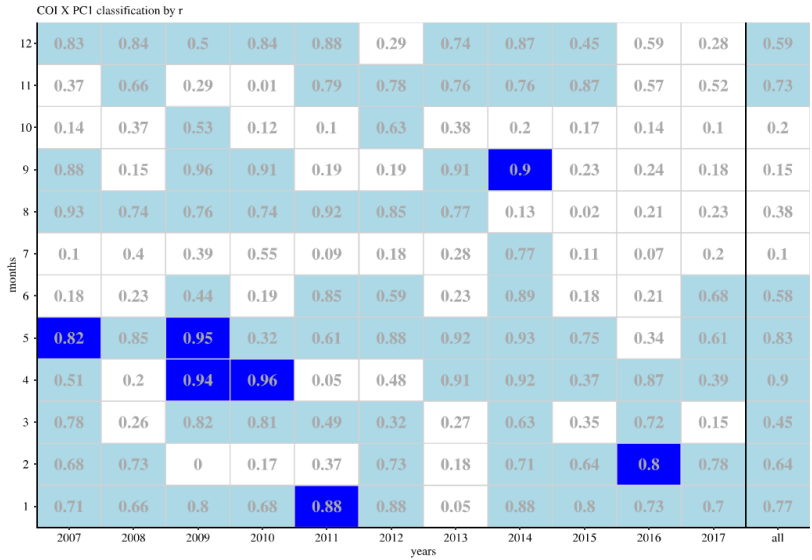
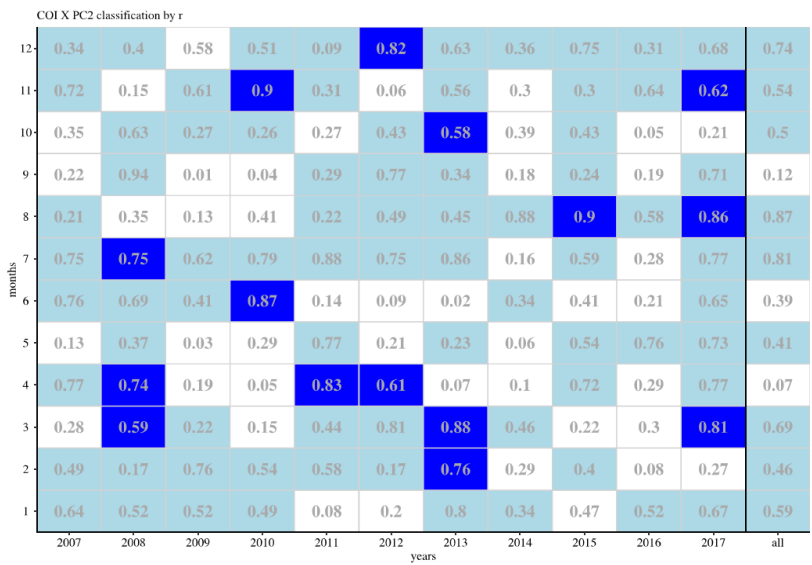
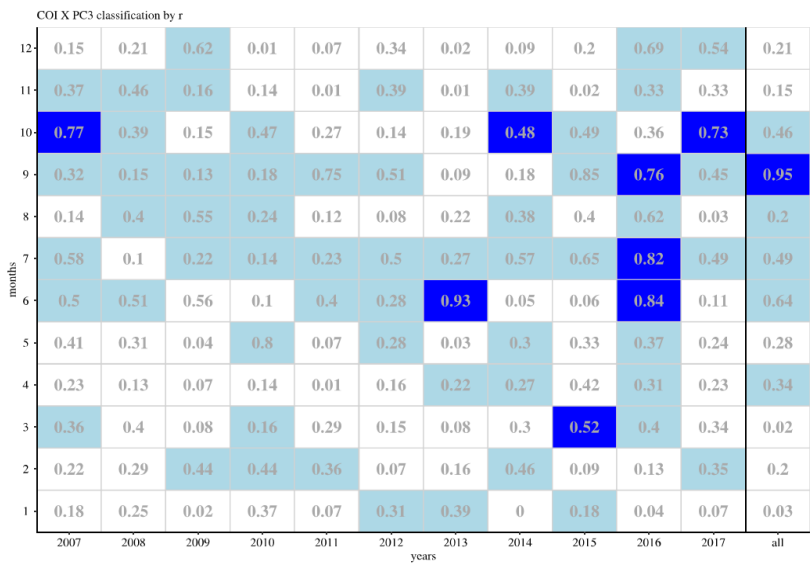

**Figure 14**. Combined classification for the X component with $Sq_{DIFl3}$ as a reference series and r is the classification parameter: correlation coefficients (numbers) between the $Sq_{DIFl3}$ and PC1 (top), PC2 (middle) and PC3 (bottom) series for different months and different years. Blue tiles mark PCs classified as Sq and light blue tiles mark PCs classified as Sq in pairs with another PC (see also Figure 15).

## COI X PC1.2 classification by r

| months | 2007 | 2008 | 2009 | 2010 | 2011 | 2012 | 2013 | 2014 | 2015 | 2016 | 2017 | all |
|---|---|---|---|---|---|---|---|---|---|---|---|---|
| 12 | 0.89 | 0.88 | 0.68 | 0.97 | 0.89 | 0.4 | 0.87 | 0.88 | 0.79 | 0.63 | 0.55 | 0.83 |
| 11 | 0.62 | 0.66 | 0.61 | 0.18 | 0.81 | 0.69 | 0.9 | 0.78 | 0.89 | 0.66 | 0.56 | 0.77 |
| 10 | 0.3 | 0.46 | 0.59 | 0.17 | 0.28 | 0.69 | 0.46 | 0.27 | 0.22 | 0.14 | 0.12 | 0.26 |
| 9 | 0.89 | 0.93 | 0.94 | 0.88 | 0.29 | 0.57 | 0.94 | 0.9 | 0.26 | 0.27 | 0.49 | 0.19 |
| 8 | 0.94 | 0.76 | 0.76 | 0.75 | 0.94 | 0.91 | 0.84 | 0.88 | 0.5 | 0.38 | 0.65 | 0.87 |
| 7 | 0.42 | 0.56 | 0.64 | 0.7 | 0.58 | 0.73 | 0.88 | 0.77 | 0.47 | 0.13 | 0.55 | 0.64 |
| 6 | 0.37 | 0.41 | 0.58 | 0.54 | 0.58 | 0.3 | 0.22 | 0.95 | 0.42 | 0.21 | 0.85 | 0.52 |
| 5 | 0.33 | 0.89 | 0.95 | 0.32 | 0.88 | 0.81 | 0.93 | 0.93 | 0.8 | 0.81 | 0.93 | 0.86 |
| 4 | 0.91 | 0.44 | 0.94 | 0.92 | 0.71 | 0.5 | 0.91 | 0.9 | 0.81 | 0.9 | 0.85 | 0.86 |
| 3 | 0.78 | 0.29 | 0.85 | 0.81 | 0.63 | 0.83 | 0.87 | 0.78 | 0.41 | 0.77 | 0.39 | 0.7 |
| 2 | 0.75 | 0.75 | 0.13 | 0.38 | 0.6 | 0.74 | 0.2 | 0.76 | 0.65 | 0.78 | 0.78 | 0.67 |
| 1 | 0.96 | 0.75 | 0.88 | 0.82 | 0.88 | 0.9 | 0.35 | 0.92 | 0.8 | 0.89 | 0.86 | 0.9 |

## COI X PC1.3 classification by r

| months | 2007 | 2008 | 2009 | 2010 | 2011 | 2012 | 2013 | 2014 | 2015 | 2016 | 2017 | all |
|---|---|---|---|---|---|---|---|---|---|---|---|---|
| 12 | 0.82 | 0.87 | 0.74 | 0.81 | 0.88 | 0.36 | 0.73 | 0.87 | 0.47 | 0.69 | 0.37 | 0.63 |
| 11 | 0.48 | 0.76 | 0.32 | 0.04 | 0.78 | 0.85 | 0.73 | 0.82 | 0.87 | 0.6 | 0.61 | 0.74 |
| 10 | 0.21 | 0.44 | 0.54 | 0.18 | 0.26 | 0.64 | 0.42 | 0.21 | 0.29 | 0.18 | 0.19 | 0.28 |
| 9 | 0.92 | 0.19 | 0.97 | 0.93 | 0.64 | 0.46 | 0.91 | 0.9 | 0.75 | 0.43 | 0.29 | 0.87 |
| 8 | 0.93 | 0.78 | 0.86 | 0.78 | 0.93 | 0.85 | 0.79 | 0.28 | 0.02 | 0.43 | 0.23 | 0.42 |
| 7 | 0.13 | 0.4 | 0.42 | 0.56 | 0.1 | 0.44 | 0.38 | 0.89 | 0.35 | 0.21 | 0.29 | 0.26 |
| 6 | 0.19 | 0.39 | 0.53 | 0.21 | 0.87 | 0.62 | 0.83 | 0.9 | 0.08 | 0.82 | 0.69 | 0.78 |
| 5 | 0.51 | 0.87 | 0.95 | 0.86 | 0.61 | 0.91 | 0.92 | 0.95 | 0.76 | 0.5 | 0.65 | 0.83 |
| 4 | 0.55 | 0.22 | 0.94 | 0.95 | 0.06 | 0.48 | 0.94 | 0.96 | 0.54 | 0.91 | 0.44 | 0.96 |
| 3 | 0.85 | 0.36 | 0.82 | 0.82 | 0.51 | 0.24 | 0.27 | 0.68 | 0.49 | 0.79 | 0.26 | 0.42 |
| 2 | 0.7 | 0.75 | 0.05 | 0.25 | 0.41 | 0.73 | 0.18 | 0.77 | 0.64 | 0.8 | 0.83 | 0.67 |
| 1 | 0.72 | 0.69 | 0.79 | 0.7 | 0.88 | 0.93 | 0.19 | 0.87 | 0.81 | 0.73 | 0.71 | 0.75 |

## COI X PC2.3 classification by r

| months | 2007 | 2008 | 2009 | 2010 | 2011 | 2012 | 2013 | 2014 | 2015 | 2016 | 2017 | all |
|---|---|---|---|---|---|---|---|---|---|---|---|---|
| 12 | 0.37 | 0.35 | 0.66 | 0.45 | 0.1 | 0.72 | 0.5 | 0.15 | 0.77 | 0.72 | 0.83 | 0.73 |
| 11 | 0.78 | 0.47 | 0.63 | 0.74 | 0.15 | 0.19 | 0.45 | 0.46 | 0.28 | 0.72 | 0.42 | 0.27 |
| 10 | 0.47 | 0.69 | 0.26 | 0.48 | 0.36 | 0.39 | 0.39 | 0.4 | 0.63 | 0.17 | 0.72 | 0.67 |
| 9 | 0.33 | 0.95 | 0.08 | 0.13 | 0.8 | 0.87 | 0.35 | 0.26 | 0.88 | 0.75 | 0.84 | 0.93 |
| 8 | 0.19 | 0.52 | 0.57 | 0.28 | 0.23 | 0.4 | 0.49 | 0.93 | 0.9 | 0.85 | 0.86 | 0.89 |
| 7 | 0.8 | 0.75 | 0.65 | 0.8 | 0.9 | 0.9 | 0.9 | 0.44 | 0.81 | 0.75 | 0.89 | 0.93 |
| 6 | 0.79 | 0.83 | 0.56 | 0.82 | 0.15 | 0.18 | 0.88 | 0.35 | 0.41 | 0.84 | 0.63 | 0.71 |
| 5 | 0.42 | 0.49 | 0.05 | 0.81 | 0.77 | 0.33 | 0.24 | 0.3 | 0.54 | 0.83 | 0.77 | 0.5 |
| 4 | 0.8 | 0.72 | 0.14 | 0.13 | 0.83 | 0.19 | 0.23 | 0.25 | 0.78 | 0.42 | 0.81 | 0.3 |
| 3 | 0.39 | 0.48 | 0.23 | 0.2 | 0.49 | 0.82 | 0.88 | 0.52 | 0.43 | 0.49 | 0.77 | 0.53 |
| 2 | 0.52 | 0.29 | 0.88 | 0.67 | 0.65 | 0.18 | 0.62 | 0.46 | 0.2 | 0.15 | 0.36 | 0.3 |
| 1 | 0.65 | 0.57 | 0.37 | 0.53 | 0.08 | 0.37 | 0.84 | 0.23 | 0.18 | 0.52 | 0.63 | 0.44 |

**Figure 15**. Same as Fig. 14 but for a pair of PCs (top – PC1+PC2, middle – PC1+PC3, bottom – PC1+PC3).

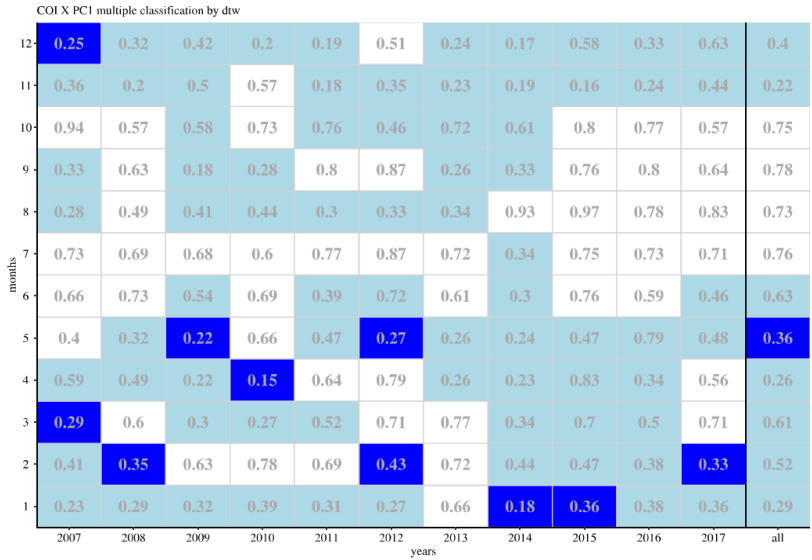
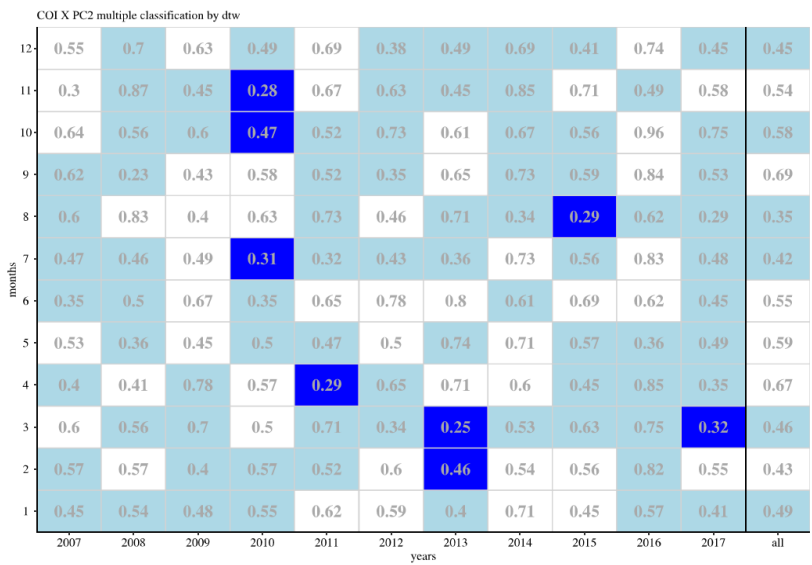
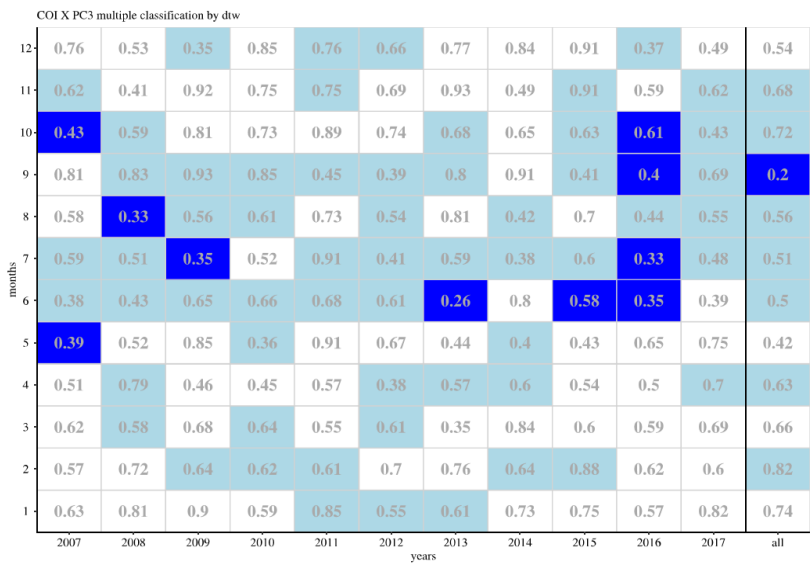

**Figure 16**. Combined classification for the X component with $Sq_{DIFl3}$ as a reference series and dtw is the classification parameter: dtw values (numbers) between the $Sq_{DIFl3}$ and PC1 (top), PC2 (middle) and PC3 (bottom) series for different months and different years. Blue tiles mark PCs classified as Sq and light blue tiles mark PCs classified as Sq in pairs with another PC (see also Figure 17).

### COI X PC1.2 multiple classification by dtw

| months | 2007 | 2008 | 2009 | 2010 | 2011 | 2012 | 2013 | 2014 | 2015 | 2016 | 2017 | all |
|---|---|---|---|---|---|---|---|---|---|---|---|---|
| 12 | 0.26 | 0.28 | 0.51 | 0.15 | 0.19 | 0.44 | 0.18 | 0.16 | 0.26 | 0.32 | 0.36 | 0.2 |
| 11 | 0.4 | 0.19 | 0.3 | 0.49 | 0.16 | 0.25 | 0.17 | 0.17 | 0.16 | 0.22 | 0.4 | 0.2 |
| 10 | 0.73 | 0.51 | 0.42 | 0.65 | 0.35 | 0.38 | 0.66 | 0.58 | 0.8 | 0.61 | 0.54 | 0.68 |
| 9 | 0.31 | 0.28 | 0.24 | 0.26 | 0.65 | 0.59 | 0.23 | 0.32 | 0.69 | 0.77 | 0.65 | 0.65 |
| 8 | 0.26 | 0.47 | 0.41 | 0.44 | 0.26 | 0.29 | 0.26 | 0.36 | 0.7 | 0.68 | 0.57 | 0.34 |
| 7 | 0.63 | 0.6 | 0.54 | 0.52 | 0.55 | 0.54 | 0.32 | 0.37 | 0.76 | 0.71 | 0.57 | 0.64 |
| 6 | 0.62 | 0.67 | 0.56 | 0.63 | 0.56 | 0.67 | 0.59 | 0.22 | 0.69 | 0.59 | 0.34 | 0.56 |
| 5 | 0.52 | 0.31 | 0.23 | 0.66 | 0.35 | 0.41 | 0.24 | 0.24 | 0.36 | 0.3 | 0.23 | 0.37 |
| 4 | 0.24 | 0.5 | 0.21 | 0.24 | 0.44 | 0.77 | 0.25 | 0.25 | 0.35 | 0.25 | 0.33 | 0.3 |
| 3 | 0.36 | 0.59 | 0.23 | 0.25 | 0.31 | 0.31 | 0.32 | 0.33 | 0.42 | 0.43 | 0.62 | 0.42 |
| 2 | 0.33 | 0.35 | 0.58 | 0.66 | 0.55 | 0.45 | 0.7 | 0.41 | 0.46 | 0.35 | 0.33 | 0.5 |
| 1 | 0.19 | 0.25 | 0.24 | 0.32 | 0.3 | 0.27 | 0.53 | 0.18 | 0.36 | 0.2 | 0.25 | 0.21 |

### COI X PC1.3 multiple classification by dtw

| months | 2007 | 2008 | 2009 | 2010 | 2011 | 2012 | 2013 | 2014 | 2015 | 2016 | 2017 | all |
|---|---|---|---|---|---|---|---|---|---|---|---|---|
| 12 | 0.26 | 0.3 | 0.29 | 0.23 | 0.18 | 0.46 | 0.21 | 0.2 | 0.56 | 0.3 | 0.63 | 0.41 |
| 11 | 0.19 | 0.26 | 0.46 | 0.53 | 0.14 | 0.27 | 0.21 | 0.2 | 0.15 | 0.22 | 0.34 | 0.19 |
| 10 | 0.89 | 0.49 | 0.57 | 0.72 | 0.83 | 0.48 | 0.6 | 0.61 | 0.81 | 0.73 | 0.57 | 0.68 |
| 9 | 0.31 | 0.65 | 0.17 | 0.24 | 0.48 | 0.6 | 0.21 | 0.33 | 0.48 | 0.7 | 0.6 | 0.34 |
| 8 | 0.27 | 0.45 | 0.35 | 0.4 | 0.3 | 0.27 | 0.36 | 0.75 | 0.97 | 0.65 | 0.83 | 0.71 |
| 7 | 0.73 | 0.68 | 0.64 | 0.58 | 0.77 | 0.53 | 0.59 | 0.29 | 0.71 | 0.69 | 0.63 | 0.67 |
| 6 | 0.65 | 0.65 | 0.53 | 0.65 | 0.38 | 0.42 | 0.32 | 0.3 | 0.59 | 0.36 | 0.44 | 0.42 |
| 5 | 0.4 | 0.32 | 0.22 | 0.37 | 0.47 | 0.28 | 0.26 | 0.2 | 0.46 | 0.59 | 0.51 | 0.45 |
| 4 | 0.57 | 0.4 | 0.21 | 0.19 | 0.58 | 0.46 | 0.21 | 0.17 | 0.5 | 0.29 | 0.51 | 0.17 |
| 3 | 0.32 | 0.58 | 0.3 | 0.22 | 0.49 | 0.56 | 0.75 | 0.34 | 0.66 | 0.44 | 0.65 | 0.52 |
| 2 | 0.42 | 0.35 | 0.62 | 0.73 | 0.69 | 0.43 | 0.71 | 0.38 | 0.45 | 0.38 | 0.35 | 0.41 |
| 1 | 0.22 | 0.29 | 0.31 | 0.38 | 0.28 | 0.21 | 0.59 | 0.18 | 0.42 | 0.35 | 0.34 | 0.28 |

### COI X PC2.3 multiple classification by dtw

| months | 2007 | 2008 | 2009 | 2010 | 2011 | 2012 | 2013 | 2014 | 2015 | 2016 | 2017 | all |
|---|---|---|---|---|---|---|---|---|---|---|---|---|
| 12 | 0.62 | 0.4 | 0.56 | 0.52 | 0.59 | 0.37 | 0.47 | 0.83 | 0.38 | 0.34 | 0.36 | 0.31 |
| 11 | 0.38 | 0.36 | 0.45 | 0.39 | 0.64 | 0.49 | 0.57 | 0.42 | 0.64 | 0.3 | 0.54 | 0.59 |
| 10 | 0.6 | 0.48 | 0.59 | 0.53 | 0.46 | 0.67 | 0.69 | 0.65 | 0.53 | 0.79 | 0.36 | 0.48 |
| 9 | 0.44 | 0.22 | 0.45 | 0.53 | 0.4 | 0.28 | 0.62 | 0.73 | 0.37 | 0.4 | 0.31 | 0.26 |
| 8 | 0.54 | 0.46 | 0.54 | 0.55 | 0.73 | 0.39 | 0.62 | 0.27 | 0.29 | 0.4 | 0.28 | 0.33 |
| 7 | 0.42 | 0.45 | 0.47 | 0.38 | 0.28 | 0.24 | 0.28 | 0.57 | 0.47 | 0.42 | 0.3 | 0.26 |
| 6 | 0.33 | 0.33 | 0.6 | 0.33 | 0.64 | 0.66 | 0.32 | 0.57 | 0.6 | 0.35 | 0.48 | 0.49 |
| 5 | 0.39 | 0.34 | 0.58 | 0.35 | 0.47 | 0.44 | 0.71 | 0.36 | 0.57 | 0.34 | 0.5 | 0.6 |
| 4 | 0.37 | 0.47 | 0.5 | 0.55 | 0.31 | 0.37 | 0.5 | 0.48 | 0.37 | 0.43 | 0.31 | 0.41 |
| 3 | 0.44 | 0.53 | 0.66 | 0.47 | 0.64 | 0.3 | 0.25 | 0.5 | 0.58 | 0.59 | 0.47 | 0.48 |
| 2 | 0.57 | 0.58 | 0.32 | 0.47 | 0.47 | 0.51 | 0.46 | 0.56 | 0.8 | 0.78 | 0.6 | 0.74 |
| 1 | 0.42 | 0.59 | 0.66 | 0.53 | 0.84 | 0.34 | 0.32 | 0.51 | 0.75 | 0.52 | 0.4 | 0.47 |

**Figure 17**. Same as Fig. 16 but for a pair of PCs (top – PC1+PC2, middle – PC1+PC3, bottom – PC1+PC3).

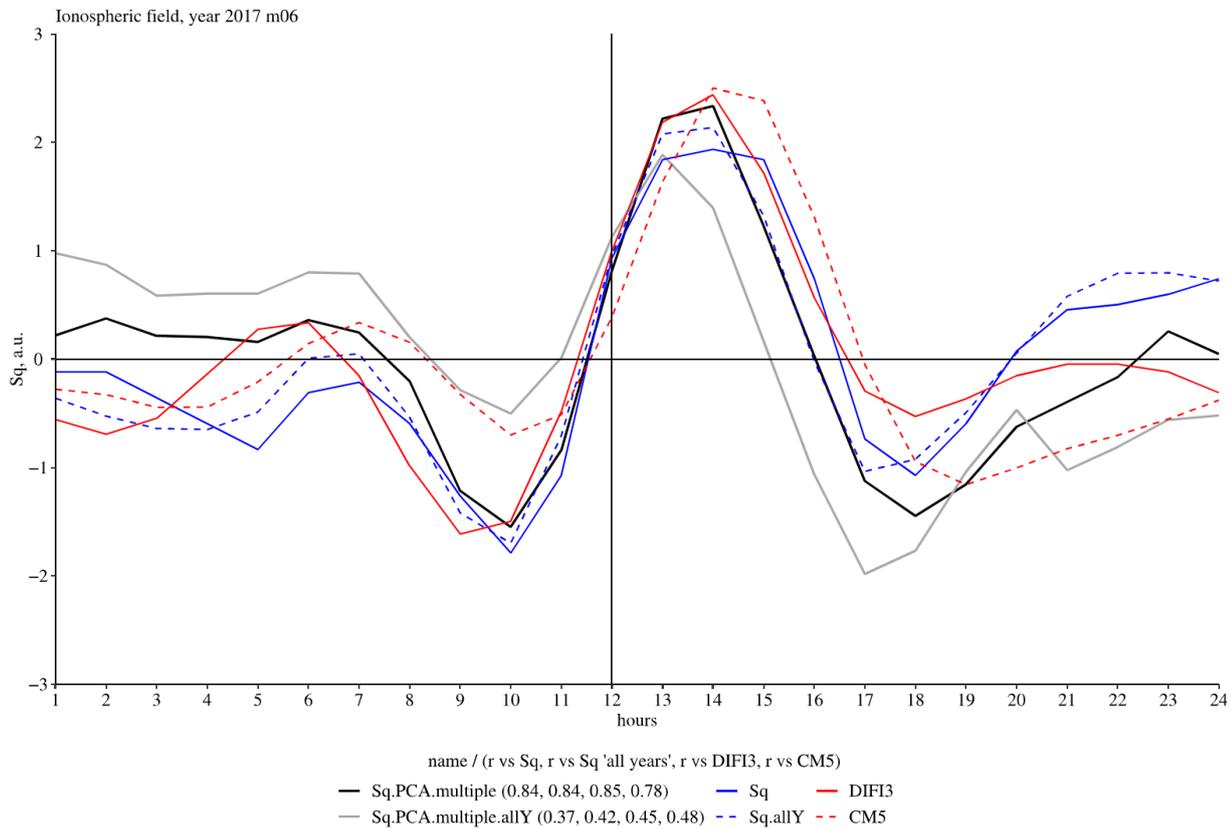

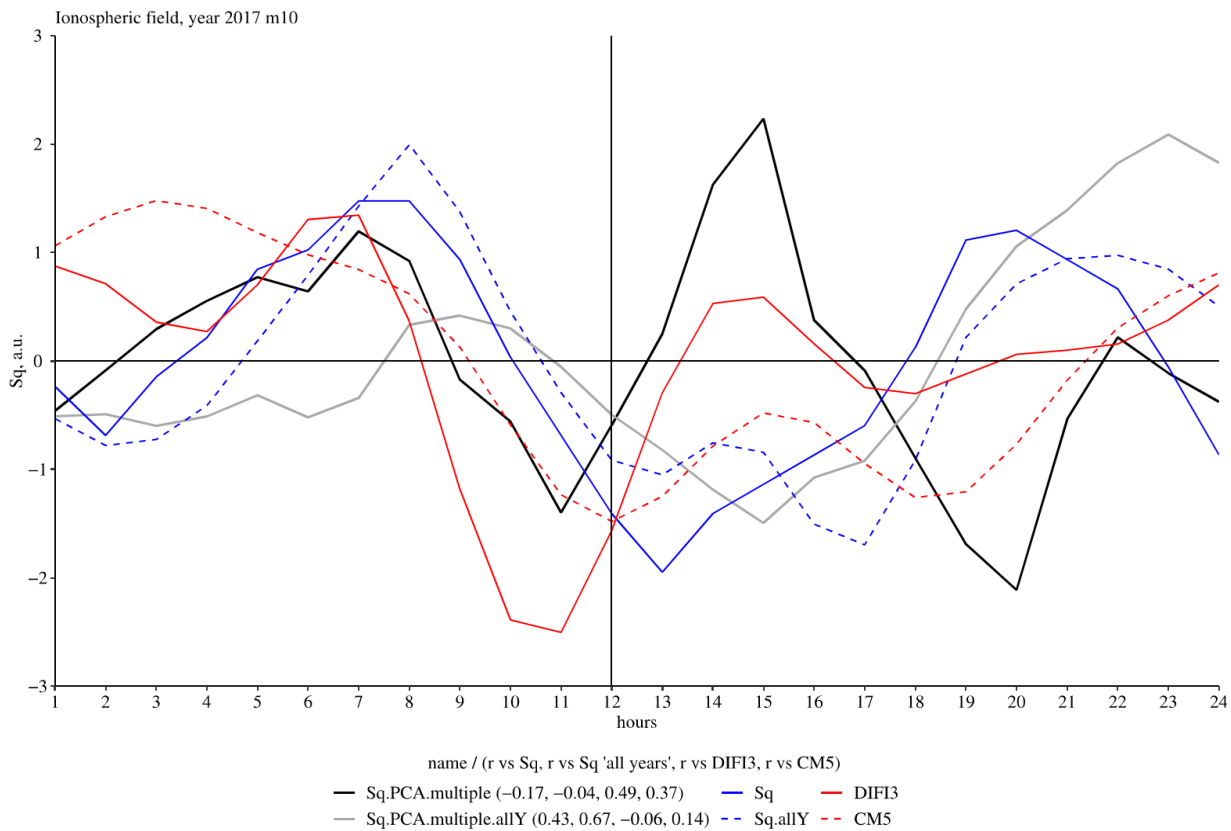

**Figure 18**. Examples of different types of Sq for the X component. Sq-type variations observed or predicted for June (top) and October (bottom) of 2017: $Sq_{PCA}$ for June or October of 2017 – black lines; $Sq_{PCA}$ for June or October of "all years" – grey lines; $Sq_{IQD}$ for June or October of 2017 – blue solid lines, $Sq_{IQD\ allY}$ for June or October of "all years" – blue dashed lines; $Sq_{DIFI3}$ for June or October – red solid line; $Sq_{CM5}$ for June or October – red dashed line. Corresponding correlation coefficients between $Sq_{PCA}$ and reference series are shown below the plots: ($r_{vs\ Sq\_IQD:indY}$, $r_{vs\ Sq\_IQD\_allY}$, $r_{vs\ Sq\_DIFI3}$, $r_{vs\ Sq\_CM5}$).

# 7. Recommendations and comments on the usage of PCA as a method to extract Sq variation from the geomagnetic filed measurements

In this work we presented a detailed analysis of the performance of the principal component analysis (PCA) as a tool to extract Sq variation from the geomagnetic field observations (X, Y and Z components) made at a mid-latitudinal station (Coimbra Magnetic Observatory, Portugal).

The studied time interval is from January 2007 to December 2017; the time resolution is 1h. The data were analyzed individually for all 12 months. The geomagnetic field components were analyzed separately.

The PCA modes were compared to Sq variation obtained for the same month using the standard approach based on the calculation of the mean daily variations using 5 international quiet days (IQD), $Sq_{IQD}$, using both the correlation and the dynamic time warping (DTW) analyses. Also, the CM5 and DIFI3 models were used to generate reference series of the ionospheric field.

Only the first three PCA modes were analyzed.

Based on the correlation or DTW analyses some PCs were classified as $Sq_{PCA}$. Two approaches were tested: only one PC can be classified as $Sq_{PCA}$ (single classification) or also a weighted sum of PCs can be classified as $Sq_{PCA}$ (combined classification).

The number of the PCs classified as Sq and their order were analyzed in relation to the component (X, Y or Z), season (mean seasonal variations through a year) and year (mean decadal variations during the 11-year solar/geomagnetic activity cycle).

*Recommendations for the use of PCA to extract Sq:*

1. It was found that for the Y and Z components the Sq variation is always filtered to the first PCA mode (PC1). Thus, PCA can be used to extract Sq variations from the observations of the Y and Z geomagnetic field components without any additional procedures.
    1.1. For the studied time interval and for the mid-latitudinal geomagnetic data we found no limitations or constraints to the usage of PCA as a tool to extract Sq variations from the Y and Z components' series.
    1.2. There are no significant differences in PC1s obtained for a certain month for an individual year and several years (11 years in our case), thus the input data set with the length of ~1 month (not necessary coinciding with a calendar month) will be sufficient to extract reliable Sq variation from the series of the Y and Z component.
2. The classification of PCs obtained for the X component is much more complicated, probably, due to the higher contribution of the geomagnetic disturbances in the variability of the X component at the middle latitudes:
    2.1. All three first PCs can be classified as Sq.
    2.2. No patterns in the classification rate of different PCs related to the season or the level of the solar/geomagnetic activity were found.
    2.3. Thus, PCA still can be used to extract Sq variation from the observations of the X component, but further analysis, for example, a comparison to a set of reference curves either obtained from the data analysis or generated using models, is always needed to classify PCs of the X component.
        2.3.1. Two types of reference series were tested: the $Sq_{IQD}$ series obtained for each month using all years of observations (11 years), and the ionospheric magnetic field modelled using the CM5 and DIFI3 models.
        2.3.2. The reference series were compared to PCs using two metrics: the correlation coefficient r and the DTW distance (dtw).
        2.3.3. In general, all reference series and both metrics performed well, however only the combination of the DIFI3 model as a reference series and the dtw metric allowed us to identify $Sq_{PCA}$ for all analyzed series.
        2.3.4. We recommend to use the DIFI-class models to generate ionospheric field.
        2.3.5. We recommend to use the dtw metric to classify $Sq_{PCA}$.
3. The reference series may be successfully used to solve a "sign ambiguity problem" for PCs that are obtained using SVD. There is no general way to solve the sign ambiguity, and comparison to a reference series can be an easy way to do so. This may apply not only to the X but also to the Y and Z series.

Finally, we summarize the main advantages and a disadvantage/constrain of the usage of the PCA method to extract Sq-type variation from the observations of the geomagnetic field components that we found:

*PCA advantages:*
1. With PCA there is no need to estimate the (relative) level of geomagnetic activity of different days of an analyzed month (or another time interval of a comparable length) to find geomagnetically quiet days (e.g., international or local quiet days). All available days of an analyzed time interval can be used.
2. It is well known that $Sq_{IQD}$ can be contaminated by the disturbance field (Yamazaki and Maute, 2017) since not all IQDs of a certain month could be quiet in the absolute sense. However, since PCA is applied to the month-long time interval, this method may allow extracting the Sq variation that has less contribution from the disturbance field. Thus, PCA allows to minimize the effect of the geomagnetic activity during individual days and to obtain the Sq variation that is more typical for the studied month.
3. The shape of the Sq variation observed at a certain location depends on the position of the geomagnetic observatory relative to the focus of the Sq current vortex. Thus, under certain circumstances, $Sq_{IQD}$ could reflect not the general conditions in the ionosphere and the upper atmosphere during a certain month but some individual features (position of the focus and the shape) of the vortex during IQDs. On the other hand, while PCA is applied to the month-long time interval, this method may allow to minimize the effect of the individual days and to obtain a more "climatological" Sq variation.
4. PCA allows the estimation of the variance fraction associated with a mode that is classified as Sq.
5. The EOF functions available for each of the PCs for each day of the analyzed time interval permit reconstructing the amplitudes of the Sq variation for each day individually allowing the assessment of its day-to-day variability.

*PCA disadvantage:*
1. The automatic classification of PCs is not always straightforward. For the Y and Z components, the Sq variations seem to be always filtered to PC1, however for the X component an additional manual or automatic classification is needed (e.g., by comparing PCs to a set of reference curves as proposed above).

The list of the abbreviations used in this paper can be found in the Supplementary Material (SM8).

The list of the datasets and software used to validate method's performance can be found in the Supplementary Material (SM9).

### CRediT author statement
Anna Morozova: Conceptualization, Formal analysis, Methodology, Supervision, Validation, Software, Investigation, Visualization, Writing - original draft.
Rania Rebbah: Software, Investigation, Data processing, Visualization, Writing - review & editing.

### Acknowledgments
AM is thankful to Dr. T. Giorgino for the development of the "dtw" R package (https://dynamictimewarping.github.io/ , http://dtw.r-forge.r-project.org/).
Funding: CITEUC is funded by the National Funds through FCT (Foundation for Science and Technology) projects UID/00611/2020 and UIDP/00611/2020.
IA is supported by FCT through the research grants UIDB/04434/2020 and UIDP/04434/2020.
This study is a contribution to the MAG-GIC project (PTDC/CTA-GEO/31744/2017), and RR is funded through this project.

### Declaration of interests
*Please **tick** the appropriate statement below (please <u>do not delete</u> either statement) and declare any financial interests/personal relationships which may affect your work in the box below.*

☒ The authors declare that they have no known competing financial interests or personal relationships that could have appeared to influence the work reported in this paper.

☐ The authors declare the following financial interests/personal relationships which may be considered as potential competing interests:

### Supplementary material *and/or* additional information: SM1-SM9

# References

Amory-Mazaudier, C (1994) On the electric current systems in the Earth's environment some historical aspects Part I: external part/ ionosphere/quiet variation, special issue on the IAGA meeting in Buenos Aires published in GEOACTA 21, 1-15.

Amory-Mazaudier, C. (2001) On the electric current systems in the Earth's environment some historical aspects. Part II: external part/ionosphere/disturbed variation from the IAGA Assembly in Hanoï Vietnam.

Amory-Mazaudier, C (2009) Electric current systems in the earth's environment J. Space Res., 8, pp.178-255, Niger.

Anad, F., Amory-Mazaudier, C., Hamoudi, M., Bourouis, S., Abtout, A. and Yizengaw, E. (2016) Sq solar variation at Medea Observatory (Algeria), from 2008 to 2011. Advances in Space Research, 58(9), pp.1682-1695.doi: 10.1016/j.asr.2016.06.029

Bhardwaj, S.K. and Rao, September, P.S. (2016) Longitudinal inequalities in Sq current system along 200-2100 E meridian. J. Ind. Geophys. Union, 20(5), pp.462-471.

Bhardwaj, S.K., Rao, P.S. and Veenadhari, B. (2015) Abnormal quiet day variations in Indian region along 75 E meridian. Earth, Planets and Space, 67(1), pp.1-15.

Björnsson, H. and Venegas, S.A.(1997) A manual for EOF and SVD analyses of climatic data. CCGCR Report, 97(1), pp.112-134.

Chapman, S. and Bartels, J., (1940) Geomagnetism. Oxford University Press, Oxford.

Chen, G.X., Xu, W.Y., Du, A.M., Wu, Y.Y., Chen, B. and Liu, X.C. (2007): Statistical characteristics of the day-to-day variability in the geomagnetic Sq field. Journal of Geophysical Research: Space Physics, 112(A6), doi:10.1029/2006JA012059.

Chulliat, A., Vigneron, P., & Hulot, G. (2016)First results from the Swarm Dedicated Ionospheric Field Inversion Chain Swarm Science Results after two years in Space 1. Geomagnetism. Earth, Planets and Space, 68(1). doi:10.1186/s40623-016-0481-6

Chulliat, A., Vigneron, P., Thébault, E., Sirol, O., & Hulot, G. (2013)Swarm SCARF dedicated ionospheric field inversion chain. Earth, Planets and Space, 65(11), 1271–1283. doi:10.5047/eps.2013.08.006

De Michelis, P., Tozzi, R. and Consolini, G. (2010) Principal components' features of mid-latitude geomagnetic daily variation. Ann. Geophys, 28, pp.2213-2226, doi:10.5194/angeo-28-2213-2010.

De Michelis, P., Tozzi, R. and Meloni, A. (2009) On the terms of geomagnetic daily variation in Antarctica. Ann. Geophys, 27, pp.2483-2490.

Ebisuzaki, W. (1997) A method to estimate the statistical significance of a correlation when the data are serially correlated, J. Clim., 10 (9), 2147-2153.

Giorgino, T., (2009) Computing and visualizing dynamic time warping alignments in R: the dtw package. Journal of statistical Software, 31(1), pp.1-24, doi: 10.18637/jss.v031.i07.

Golovkov V.P., Zvereva T.I. (1998) Expansion of Geomagnetic Variations within a Year in Natural Orthogonal Components, Geomagn. Aeron., 38, 368-372

Golovkov, V. P., and T. I. Zvereva (2000), The space-time pattern of midlatitude geomagnetic variations, Geomagn. Aeron., 40, 84–92.

Golovkov, V. P., N. E. Papitashvili, Y. S. Tyupkin, and E. P. Kharin (1978), Separation of geomagnetic field variations into quiet and disturbed components by the method of natural orthogonal components, Geomagn. Aeron., 18, 342–344.

Golovkov, V. P., V. O. Papitashvili, and N. E. Papitashvili (1989), Automatic calculation of K indices using the method of natural orthogonal components, Geomagn. Aeron., 29, 514–517.

Haines, G.V. and Torta, J.M. (1994): Determination of equivalent current sources from spherical cap harmonic models of geomagnetic field variations. Geophysical Journal International, 118(3), pp.499-514.

Hannachi, A., Jolliffe, I.T. and Stephenson, D.B.(2007) Empirical orthogonal functions and related techniques in atmospheric science: A review. International Journal of Climatology: A Journal of the Royal Meteorological Society, 27(9), pp.1119-1152.doi: 10.1002/joc.1499.

Maslova, I., Kokoszka, P., Sojka, J. and Zhu, L. (2010): Estimation of Sq variation by means of multiresolution and principal component analyses. Journal of Atmospheric and Solar-Terrestrial Physics, 72, 7-8, 625-632.doi: 10.1016/j.jastp.2010.02.005.

Menvielle, M., (1981) About the scalings of K indicesfrom IAGA News, 20, p.110-111.

Morozova, A.L., Rebbah, R., and Ribeiro, P., (2021a). Datasets of the solar quiet (Sq) and solar disturbed (SD) variations of the geomagnetic field at the Coimbra Magnetic Observatory (COI) obtained by different methods, Data in Brief, v. 37C, 107174, doi:10.1016/j.dib.2021.107174.

Morozova, A.L., Rebbah, R., and Ribeiro, P., (2021b). Datasets of the solar quiet (Sq) and solar disturbed (SD) variations of the geomagnetic field at a midlatitudinal station in Europe obtained by different methods, Mendeley Data, V1, doi: 10.17632/jcmdrm5f5x.1, http://dx.doi.org/10.17632/jcmdrm5f5x.1

Morozova, A.L., Ribeiro, P. and Pais, M.A., (2014) Correction of artificial jumps in the historical geomagnetic measurements of Coimbra Observatory, Portugal. Annales Geophysicae, Vol. 32, No. 1, pp. 19-40, doi:10.5194/angeo-32-19-2014.

Morozova, A.L., Ribeiro, P. and Pais, M.A., (2021c) Homogenization of the historical series from the Coimbra Magnetic Observatory, Portugal. Earth System Science Data, 13, 809–825, doi: 10.5194/essd-13-809-2021.

Piersanti, M., Alberti, T., Bemporad, A., Berrilli, F., Bruno, R., Capparelli, V., Carbone, V., Cesaroni, C., Consolini, G., Cristaldi, A. and Del Corpo, A. (2017): Comprehensive analysis of the geoeffective solar event of 21 June 2015: Effects on the magnetosphere, plasmasphere, and ionosphere systems. Solar Physics, 292(11), p.169.doi:10.1007/978-94-024-1570-4_12.

Rangarajan, G.K. and Murty, A.V.S., (1980) Scaling K-indices without subjectivityFrom IAGA news, 19, 112-118.

Sabaka, T. J., Olsen, N., & Langel, R. A. (2002). A comprehensive model of the quiet-time, near-Earth magnetic field: Phase 3. Geophysical Journal International, 151(1), 32–68. doi:10.1046/j.1365-246X.2002.01774.x

Sabaka, T.J., Tøffner-Clausen, L., Olsen, N. and Finlay, C.C., (2020) CM6: a comprehensive geomagnetic field model derived from both CHAMP and Swarm satellite observations. Earth, Planets and Space, 72, pp.1-24.doi: 10.1186/s40623-020-01210-5.

Shlens, J., (2009) A Tutorial on Principal Component Analysis Center for Neural Science. New York University New York City, NY, pp.10003-660

Stening, R., Reztsova, T. and Minh, L.H. (2005): Day-to-day changes in the latitudes of the foci of the Sq current system and their relation to equatorial electrojet strength. Journal of Geophysical Research: Space Physics, 110(A10), A10308, doi:10.1029/2005JA011219.

Stening, R.J. (2008): The shape of the Sq current system. Annales Geophysicae (Vol. 26, No. 7, pp. 1767-1775).

Takeda, M. (1982). Three dimensional ionospheric currents and field aligned currents generated by asymmetric dynamo action in the ionosphere. Journal of Atmospheric and Terrestrial Physics, 44(2), 187–193. doi:10.1016/0021-9169(82)90122-2.

Thébault, E., Vigneron, P., Langlais, B., & Hulot, G. (2016). A Swarm lithospheric magnetic field model to SH degree 80 Swarm Science Results after two years in Space 1. Geomagnetism. Earth, Planets and Space, 68(1). doi:10.1186/s40623-016-0510-5

Wu, Y.Y., Xu, W.Y., Chen, G.X., Chen, B. and Liu, X.C., (2007). The Evolution Characteristics of Geomagnetic Disturbances During Geomagnetic Storm. Chinese Journal of Geophysics, 50(1), pp.1-11.

Xu, W.Y. and Kamide, Y. (2004): Decomposition of daily geomagnetic variations by using method of natural orthogonal component. Journal of Geophysical Research: Space Physics, 109(A5).doi: 10.1029/2003JA010216.

Yamazaki, Y. and Maute, A. (2017): Sq and EEJ—A review on the daily variation of the geomagnetic field caused by ionospheric dynamo currents. Space Science Reviews, 206(1-4), pp.299-405. doi: 10.1007/s11214-016-0282-z.

# Principal component analysis as a tool to extract Sq variation from the geomagnetic field observations: conditions of applicability

Anna Morozova, Rania Rebbah

## Supplementary Material, file SM1

### Solar quiet daily variations of the geomagnetic field

The Sq variation of the geomagnetic field results from an electrical current system in the ionospheric E dynamo region. This system consists of two vortices quasi-symmetric to the equator with the anti-clockwise (clockwise) electrical currents in the sunlit Northern (Southern) Hemisphere with foci located in the middle latitudes near 30-40º depending on the longitudinal sector and the hemisphere. Near the equator, they are connected to the equatorial electrojet, and in the high latitudes, they are affected by the current systems of the polar ionosphere. As the day progresses, the position of these vortices on the globe moves westward following the Sun. Thus, for any given location on the planet, the geometry of the system changes along the day returning to a similar condition after one day.

The character of the ground measured Sq variation of the geomagnetic field components X, Y and Z depends on the position of a geomagnetic observatory relative to the vortex. The change of the sign of Sq X takes place around the foci latitudes. The sign of Sq Y and Sq Z changes near the magnetic equator (see Chapman and Bartels, 1940; Amory-Mazaudier, 1994, 2001 and 2009; Anad et al., 2016; Yamazaki and Maute, 2017). The Sq X and Sq Z variations are symmetric around the local noon, while Sq Y is anti-symmetric. In the real ionosphere, the shape of the current vortex can be far from the ideal circle or oval: the vortex can, e.g., be tilted (resulting in a shift of the daily minimum of Sq X to the afternoon hours, see Amory-Mazaudier, 1994, 2001 and 2009; Anad et al., 2016), stretched or compressed. The shape of the vortex affects mostly the Sq X variation, whereas the shapes of the Sq Y and Sq Z variations are almost constant from day to day.

## References

Chapman, S. and Bartels, J., (1940) Geomagnetism. Oxford University Press, Oxford.

Amory-Mazaudier, C (1994) On the electric current systems in the Earth's environment some historical aspects Part I: external part/ ionosphere/quiet variation, special issue on the IAGA meeting in Buenos Aires published in GEOACTA 21, 1-15.

Amory-Mazaudier, C. (2001) On the electric current systems in the Earth's environment some historical aspects. Part II: external part/ionosphere/disturbed variation from the IAGA Assembly in Hanoï Vietnam.

Amory-Mazaudier, C (2009) Electric current systems in the earth's environment J. Space Res., 8, pp.178-255, Niger.

Anad, F., Amory-Mazaudier, C., Hamoudi, M., Bourouis, S., Abtout, A. and Yizengaw, E. (2016) Sq solar variation at Medea Observatory (Algeria), from 2008 to 2011. Advances in Space Research, 58(9), pp.1682-1695.doi: 10.1016/j.asr.2016.06.029

Yamazaki, Y. and Maute, A. (2017): Sq and EEJ—A review on the daily variation of the geomagnetic field caused by ionospheric dynamo currents. Space Science Reviews, 206(1-4), pp.299-405.doi: 10.1007/s11214-016-0282-z.

Principal component analysis as a tool to extract Sq variation from the geomagnetic field observations: conditions of applicability

Anna Morozova, Rania Rebbah

Supplementary Material, file SM2

## Description of the models used to simulate the ionospheric part of the geomagnetic field

Detailed descriptions of the models can be found in Sabaka et al. (2002) for the CM5 model, and Chulliat et al. (2013, 2016) and Thébault et al. (2016) for the DIFI3 model, and short summaries are presented below.

### CM5 model

CM5 is one of the versions of the so-called Comprehensive Models, developed to parametrize all the major near-Earth magnetic field sources: the core and the lithosphere fields, the M2 tidal component, the primary and induced magnetospheric fields, and the primary and induced ionospheric fields, all for different components of the magnetic field vector. They are developed by NASA/GSFC and the Danish Technical University. The details can be found in Sabaka et al., 2002.

The model can be run online at https://ccmc.gsfc.nasa.gov/models/modelinfo.php?model=CM5.

CM5 was developed from the pre-Swarm satellite data (CHAMP, Oersted and SAC-CI) and observatory data from August 2000 to January 2013. Currently, a new version, CM6 (Sabaka et al., 2020), is available.

In our work, the CM5 model outputs for the primary and induced ionospheric field were summed to obtain a single reference series for the Sq-type variation.

### DIFI3 model

The Dedicated Ionospheric Field Inversion (DIFI) model is a time-varying, spherical harmonic representation of the quiet-time Sq and the equatorial electrojet (EEJ) fields between ± 55º quasi-dipole latitudes. It is derived from a combination of Swarm satellite and magnetic observatory data: the 0501 L1b Swarm data and observatory data between December 1, 2013, and January 29, 2017. Time variations are represented by Fourier series with periods of 24h, 12h, 8h and 6h, modulated by annual and semi-annual periodicities. The spherical harmonic expansion goes to degree 60 and order 12 in geomagnetic dipole coordinates. Solar activity (represented by the F10.7 index) dependence is also included in the model. Conductivity models of oceans and continents have been used to separate primary and induced magnetic fields (Chulliat et al., 2013, 2016; Thébault et al., 2016).

The DIFI model is developed by CIRES in collaboration with the Institut de Physique du Globe de Paris (IPGP) through Swarm's Satellite Constellation Application and Research Facility (SCARF), a project funded by the European Space Agency (ESA). DIFI is an official level 2 product of the Swarm mission. DIFI models and other Swarm level 2 products are also available at https://earth.esa.int/web/guest/swarm/data-access. The model can be run online at http://geomag.colorado.edu/difi-calculator.

### References

Chulliat, A., Vigneron, P., Thébault, E., Sirol, O., & Hulot, G. (2013)Swarm SCARF dedicated ionospheric field inversion chain. Earth, Planets and Space, 65(11), 1271–1283. doi:10.5047/eps.2013.08.006

Chulliat, A., Vigneron, P., & Hulot, G. (2016)First results from the Swarm Dedicated Ionospheric Field Inversion Chain Swarm Science Results after two years in Space 1. Geomagnetism. Earth, Planets and Space, 68(1). doi:10.1186/s40623-016-0481-6

Sabaka, T. J., Olsen, N., & Langel, R. A. (2002). A comprehensive model of the quiet-time, near-Earth magnetic field: Phase 3. Geophysical Journal International, 151(1), 32–68. doi:10.1046/j.1365-246X.2002.01774.x

Sabaka, T.J., Tøffner-Clausen, L., Olsen, N. and Finlay, C.C., (2020) CM6: a comprehensive geomagnetic field model derived from both CHAMP and Swarm satellite observations. Earth, Planets and Space, 72, pp.1-24.doi: 10.1186/s40623-020-01210-5.

Thébault, E., Vigneron, P., Langlais, B., & Hulot, G. (2016). A Swarm lithospheric magnetic field model to SH degree 80 Swarm Science Results after two years in Space 1. Geomagnetism. Earth, Planets and Space, 68(1). doi:10.1186/s40623-016-0510-5

# Principal component analysis as a tool to extract Sq variation from the geomagnetic field observations: conditions of applicability

Anna Morozova, Rania Rebbah

Supplementary Material, file SM**3**

Time variations of the solar and geomagnetic activity indices

Figures S3.1-S3.4

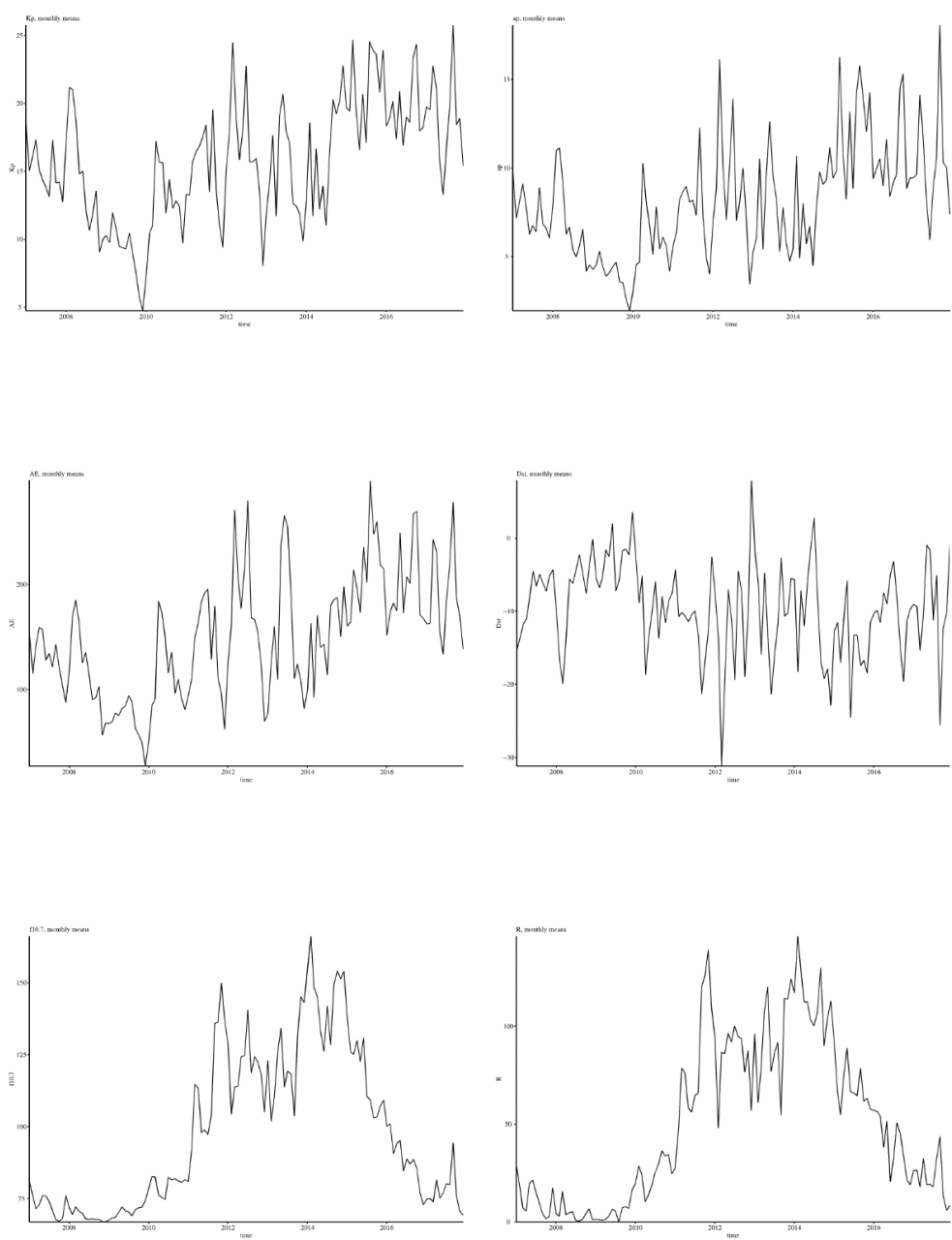

**Figure S3.1.** Time variations of the monthly mean values of the geomagnetic (top left – Kp, top right – ap, middle left – AE, middle right - Dst) and solar (bottom left – F10.7, bottom right - R) activity indices for 2007-2017.

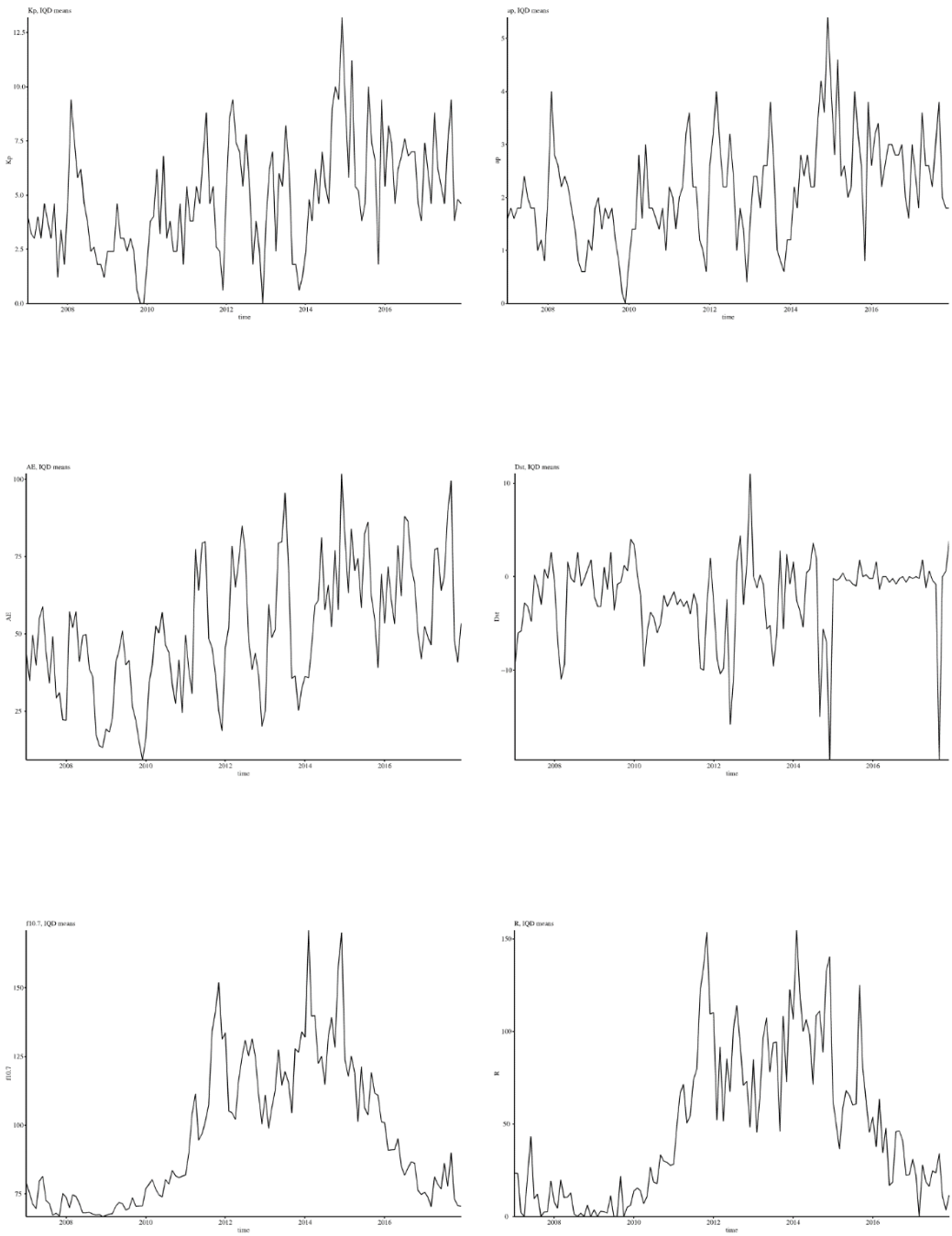

**Figure S3.2**. Same as Fig. S3.1 but for monthly IQD means.

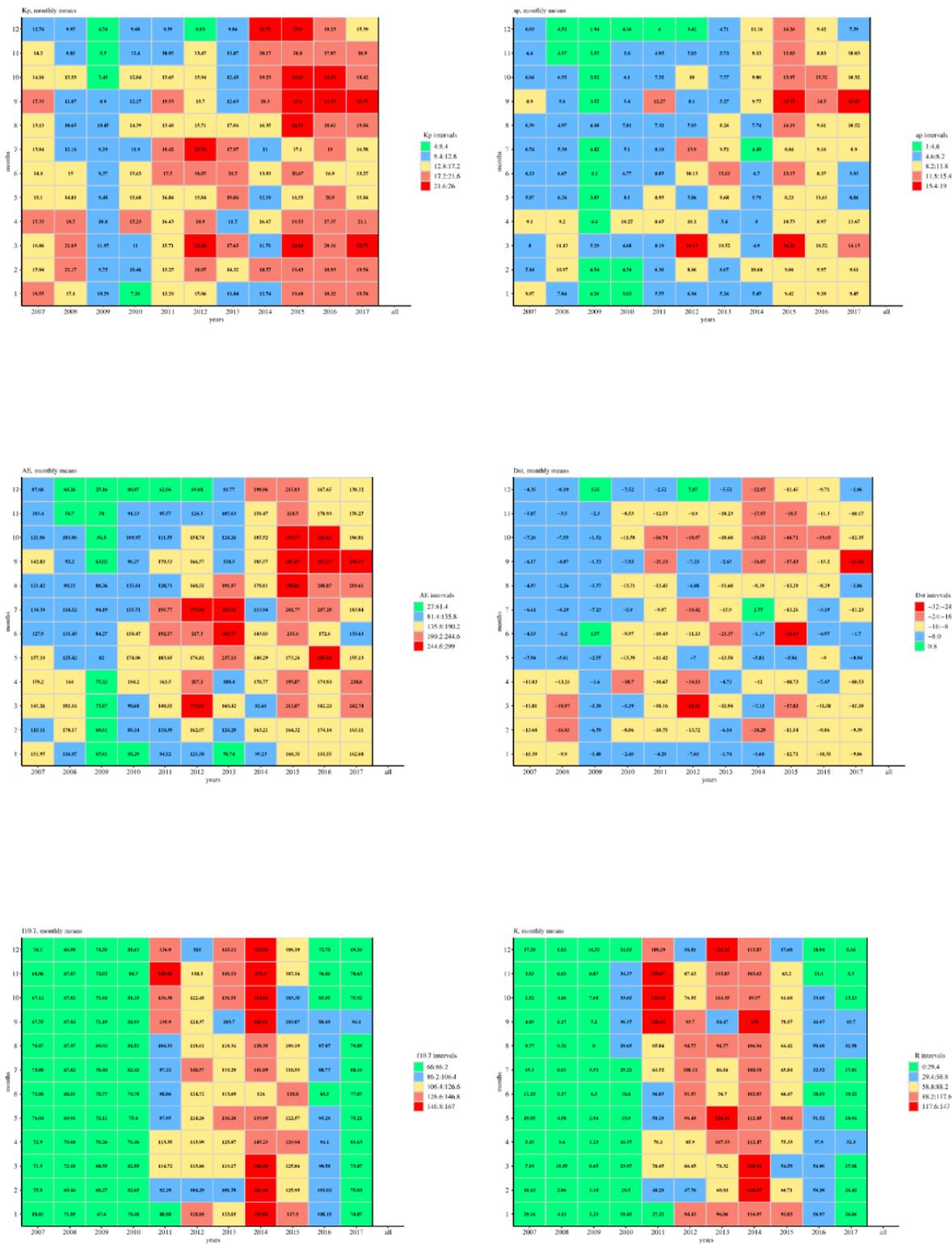

**Figure S3.3**. Tileplots with the monthly mean values of the geomagnetic (top left – Kp, top right – ap, middle left – AE, middle right - Dst) and solar (bottom left – F10.7, bottom right - R) activity indices for each month of 2007-2017 and averaged for all 11 years.

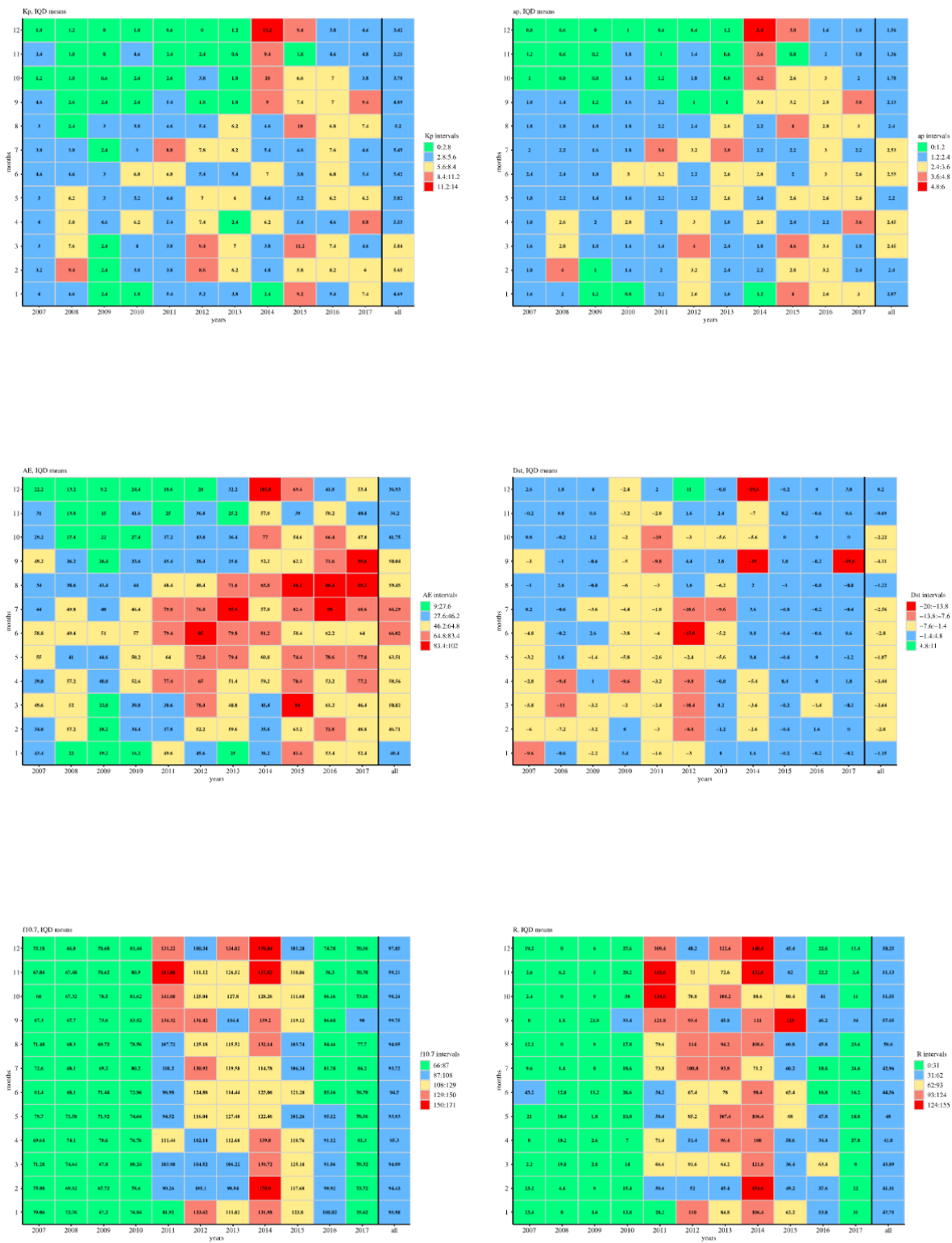

**Figure S3.4**. Same as Fig. S3.3 but for monthly IQD means.

Principal component analysis as a tool to extract Sq variation from the geomagnetic field observations: conditions of applicability

Anna Morozova, Rania Rebbah

Supplementary Material, file SM4

**Relations between dtw and r values**

**Table S4.1 and Figure S4.1**

**Table S4.1.** Correspondence between the correlation coefficients and the *dtw* parameters obtained for all sets of series studied in this work (PCs vs RS). RS = $Sq_{IQD\,allY}$, $Sq_{CM5}$ or $Sq_{DIFI3}$.

| r | dtw (mean ± SD) | | | |
|---|---|---|---|---|
| | $Sq_{IQD\,allY}$ | $Sq_{CM5}$ | $Sq_{DIFI3}$ | Mean |
| 0.3 | 0.66 ± 0.04 | 0.64 ± 0.03 | 0.69 ± 0.02 | 0.67 ± 0.03 |
| 0.5 | 0.52 ± 0.05 | 0.51 ± 0.02 | 0.59 ± 0.01 | 0.55 ± 0.04 |
| 0.6 | 0.43 ± 0.05 | 0.44 ± 0.02 | 0.52 ± 0.02 | 0.47 ± 0.05 |
| 0.75 | 0.29 ± 0.04 | 0.3 ± 0.02 | 0.39 ± 0.03 | 0.34 ± 0.05 |
| 0.9 | 0.13 ± 0.02 | 0.14 ± 0.01 | 0.19 ± 0.03 | 0.16 ± 0.03 |

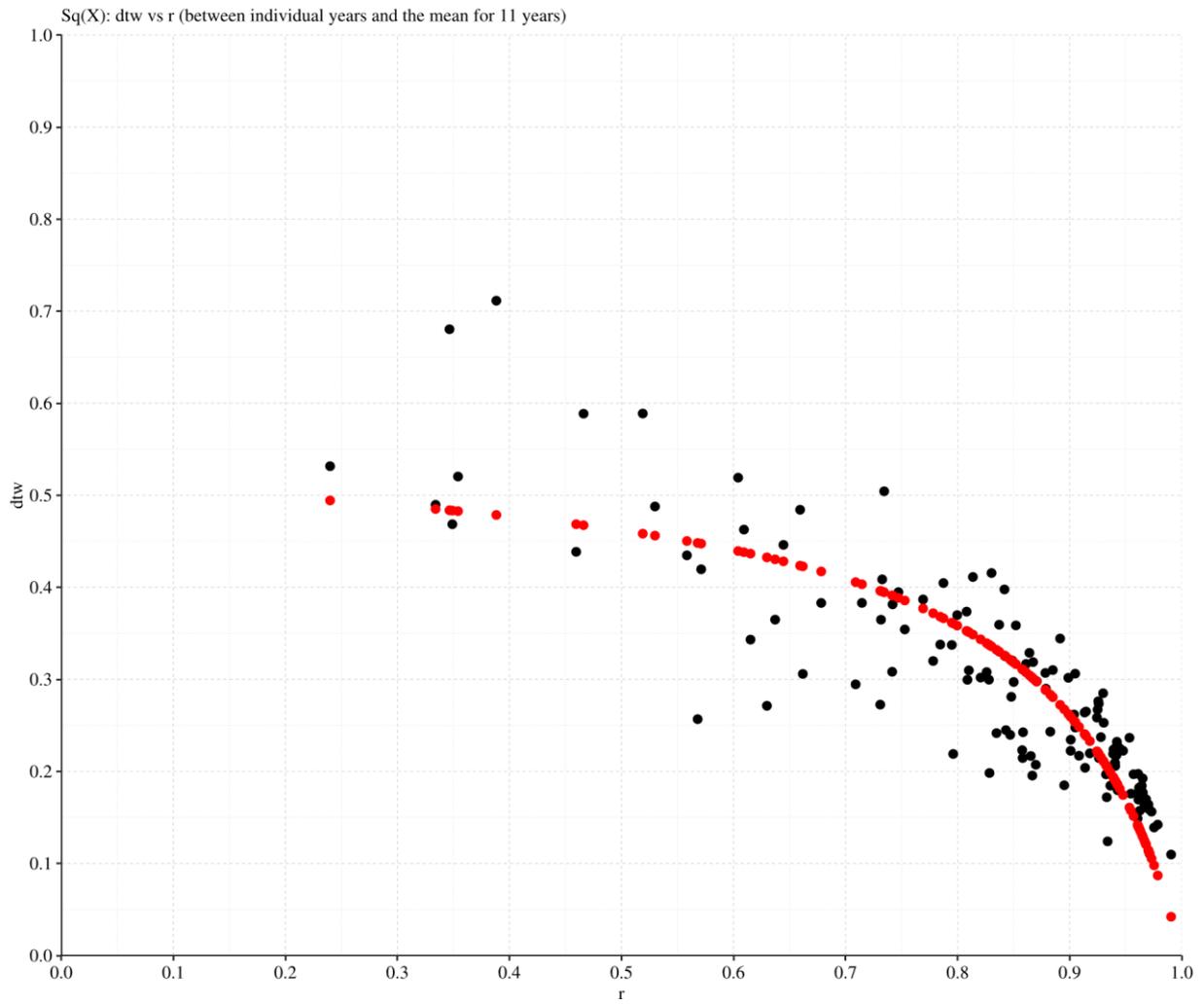

**Figure S4.1**. Comparison of the correlation coefficients *r* and *dtw* values obtained for pairs of the Sq$_{IQD}$ and Sq$_{IQD\ allY}$ series (black dots). The fit *dtw* = A(1-*r*)/(B+(1-*r*)) is shown by red dots.

Principal component analysis as a tool to extract Sq variation from the geomagnetic field observations: conditions of applicability

Anna Morozova, Rania Rebbah

Supplementary Material, file SM5

**Classification of PCs for the Y and Z components**

**Figures S5.1-S5.4**

**COI Y PC1 classification by r**

| months \ years | 2007 | 2008 | 2009 | 2010 | 2011 | 2012 | 2013 | 2014 | 2015 | 2016 | 2017 | all |
|---|---|---|---|---|---|---|---|---|---|---|---|---|
| 12 | 0.95 | 0.97 | 0.98 | 0.99 | 0.99 | 0.98 | 0.97 | 0.97 | 0.9 | 0.88 | 0.9 | 0.97 |
| 11 | 0.96 | 0.97 | 0.98 | 0.97 | 0.99 | 0.97 | 0.98 | 0.99 | 0.92 | 0.98 | 0.84 | 0.98 |
| 10 | 0.97 | 0.99 | 0.98 | 0.91 | 0.97 | 0.98 | 0.98 | 0.98 | 0.91 | 0.91 | 0.96 | 0.97 |
| 9 | 0.99 | 0.98 | 0.99 | 1 | 0.98 | 0.94 | 0.96 | 0.99 | 0.97 | 0.97 | 0.97 | 0.99 |
| 8 | 0.98 | 0.99 | 0.99 | 0.99 | 0.99 | 0.97 | 0.98 | 0.99 | 0.99 | 0.98 | 0.98 | 0.99 |
| 7 | 0.99 | 1 | 0.98 | 0.99 | 0.99 | 0.99 | 0.98 | 1 | 0.98 | 0.98 | 1 | 1 |
| 6 | 0.99 | 0.99 | 1 | 0.99 | 0.98 | 0.99 | 0.99 | 0.99 | 0.99 | 0.95 | 0.98 | 0.99 |
| 5 | 0.98 | 0.99 | 0.99 | 0.98 | 0.99 | 0.99 | 0.98 | 0.99 | 0.99 | 0.98 | 0.99 | 0.99 |
| 4 | 0.98 | 0.97 | 0.99 | 0.97 | 0.99 | 0.96 | 0.99 | 0.98 | 0.99 | 0.95 | 0.98 | 0.99 |
| 3 | 0.97 | 0.95 | 0.97 | 0.99 | 0.93 | 0.97 | 0.98 | 1 | 0.96 | 0.95 | 0.95 | 0.98 |
| 2 | 0.94 | 0.91 | 0.88 | 0.92 | 0.92 | 0.98 | 0.97 | 0.92 | 0.97 | 0.95 | 0.9 | 0.96 |
| 1 | 0.88 | 0.92 | 0.95 | 0.95 | 0.98 | 0.97 | 0.96 | 0.96 | 0.93 | 0.93 | 0.92 | 0.97 |

**COI Y PC2 classification by r**

| months \ years | 2007 | 2008 | 2009 | 2010 | 2011 | 2012 | 2013 | 2014 | 2015 | 2016 | 2017 | all |
|---|---|---|---|---|---|---|---|---|---|---|---|---|
| 12 | 0.24 | 0.12 | 0.04 | 0.08 | 0.05 | 0.07 | 0.11 | 0.13 | 0.23 | 0.35 | 0.22 | 0.13 |
| 11 | 0.24 | 0.19 | 0.13 | 0.19 | 0.07 | 0.1 | 0.08 | 0.01 | 0.34 | 0.13 | 0.48 | 0.15 |
| 10 | 0.21 | 0.05 | 0.16 | 0.39 | 0.14 | 0.01 | 0.09 | 0.11 | 0.33 | 0.38 | 0.22 | 0.19 |
| 9 | 0.14 | 0.15 | 0.11 | 0.04 | 0.19 | 0.16 | 0.23 | 0 | 0.12 | 0.14 | 0.03 | 0.11 |
| 8 | 0.11 | 0.08 | 0.08 | 0.12 | 0.04 | 0.19 | 0.17 | 0.12 | 0.06 | 0.17 | 0.17 | 0.01 |
| 7 | 0.09 | 0.07 | 0.11 | 0.02 | 0.12 | 0.14 | 0.1 | 0.02 | 0.18 | 0.13 | 0.02 | 0.03 |
| 6 | 0.11 | 0.1 | 0.04 | 0.08 | 0.12 | 0.04 | 0.07 | 0.02 | 0.01 | 0.3 | 0.05 | 0.02 |
| 5 | 0.16 | 0.09 | 0.1 | 0.16 | 0.03 | 0.09 | 0.1 | 0.16 | 0.08 | 0.18 | 0.09 | 0.08 |
| 4 | 0.2 | 0.11 | 0.1 | 0.2 | 0.12 | 0.06 | 0.14 | 0.15 | 0.12 | 0.14 | 0.18 | 0.14 |
| 3 | 0.2 | 0.27 | 0.14 | 0.08 | 0.28 | 0.18 | 0.16 | 0.02 | 0.13 | 0.14 | 0.26 | 0.13 |
| 2 | 0.17 | 0.4 | 0.16 | 0.36 | 0.35 | 0.09 | 0.17 | 0.16 | 0.2 | 0.11 | 0.4 | 0.03 |
| 1 | 0.41 | 0.06 | 0.05 | 0.26 | 0.13 | 0.11 | 0.21 | 0.19 | 0.37 | 0.23 | 0.29 | 0.19 |

**COI Y PC3 classification by r**

| months \ years | 2007 | 2008 | 2009 | 2010 | 2011 | 2012 | 2013 | 2014 | 2015 | 2016 | 2017 | all |
|---|---|---|---|---|---|---|---|---|---|---|---|---|
| 12 | 0.02 | 0.15 | 0.08 | 0.01 | 0.02 | 0.07 | 0.09 | 0.16 | 0.3 | 0.09 | 0.31 | 0.16 |
| 11 | 0.09 | 0.09 | 0.13 | 0.05 | 0.1 | 0.18 | 0.15 | 0.07 | 0.17 | 0.04 | 0.17 | 0.04 |
| 10 | 0.12 | 0.06 | 0.01 | 0.03 | 0.15 | 0.2 | 0.14 | 0.18 | 0.2 | 0.02 | 0.07 | 0.08 |
| 9 | 0.05 | 0.09 | 0.06 | 0.06 | 0.09 | 0.3 | 0.1 | 0.02 | 0.16 | 0.17 | 0.06 | 0.06 |
| 8 | 0.14 | 0.1 | 0.07 | 0.01 | 0.08 | 0.11 | 0.01 | 0.05 | 0.09 | 0.01 | 0 | 0.1 |
| 7 | 0.03 | 0.03 | 0.02 | 0.06 | 0.1 | 0.02 | 0.14 | 0.01 | 0.01 | 0.11 | 0.08 | 0.09 |
| 6 | 0.05 | 0.06 | 0.08 | 0.01 | 0.07 | 0.01 | 0.07 | 0.14 | 0.12 | 0.04 | 0.15 | 0.12 |
| 5 | 0.02 | 0.05 | 0.03 | 0.09 | 0.07 | 0.09 | 0.09 | 0.02 | 0.13 | 0.01 | 0.07 | 0.08 |
| 4 | 0.01 | 0.01 | 0.03 | 0.08 | 0.06 | 0.26 | 0.03 | 0.05 | 0.03 | 0.22 | 0.07 | 0.07 |
| 3 | 0.07 | 0.02 | 0.05 | 0.06 | 0.19 | 0.03 | 0.07 | 0.06 | 0.21 | 0.21 | 0.02 | 0.12 |
| 2 | 0.24 | 0.06 | 0.09 | 0.12 | 0.07 | 0.04 | 0.06 | 0.09 | 0.11 | 0.29 | 0.11 | 0.12 |
| 1 | 0.14 | 0.36 | 0.24 | 0.06 | 0.03 | 0.16 | 0.08 | 0.18 | 0.01 | 0.27 | 0.19 | 0.08 |

**Figure S5.1**. Single classification for the Y component with $Sq_{IQD}$ as the reference series and *r* is the classification metric: correlation coefficients (numbers) between $Sq_{IQD}$ and PC1 (top), PC2 (middle) and PC3 (bottom) series for different months (Y-axis) and different years (X-axis). Blue tiles mark PCs classified as Sq (single classification).

## COI Z PC1 classification by r

| months \ years | 2007 | 2008 | 2009 | 2010 | 2011 | 2012 | 2013 | 2014 | 2015 | 2016 | 2017 | all |
|---|---|---|---|---|---|---|---|---|---|---|---|---|
| 12 | 0.94 | 0.97 | 0.97 | 0.93 | 0.99 | 0.99 | 0.96 | 0.95 | 0.81 | 0.87 | 0.89 | 0.96 |
| 11 | 0.97 | 0.98 | 0.97 | 0.97 | 1 | 0.97 | 0.99 | 0.99 | 0.96 | 0.96 | 0.95 | 0.99 |
| 10 | 0.98 | 0.99 | 0.99 | 0.95 | 0.99 | 0.99 | 0.99 | 0.97 | 0.96 | 0.97 | 0.98 | 0.99 |
| 9 | 0.96 | 0.97 | 0.99 | 0.99 | 0.95 | 0.96 | 0.97 | 0.97 | 0.97 | 0.97 | 0.92 | 0.98 |
| 8 | 0.97 | 0.99 | 0.99 | 0.99 | 0.99 | 0.98 | 0.98 | 0.97 | 0.97 | 0.96 | 0.99 | 0.99 |
| 7 | 0.99 | 0.99 | 0.98 | 0.99 | 0.98 | 0.95 | 0.98 | 1 | 0.98 | 0.98 | 0.99 | 0.99 |
| 6 | 1 | 0.99 | 0.99 | 0.99 | 0.99 | 0.99 | 0.95 | 0.99 | 0.98 | 0.98 | 0.97 | 0.99 |
| 5 | 0.98 | 0.99 | 0.99 | 0.97 | 0.99 | 0.99 | 0.98 | 0.99 | 0.99 | 0.98 | 0.99 | 0.99 |
| 4 | 0.98 | 0.98 | 1 | 0.98 | 0.99 | 0.97 | 0.99 | 0.99 | 1 | 0.96 | 0.98 | 0.99 |
| 3 | 0.99 | 0.95 | 0.99 | 0.99 | 0.97 | 0.98 | 0.96 | 1 | 0.97 | 0.96 | 0.92 | 0.98 |
| 2 | 0.97 | 0.94 | 0.91 | 0.98 | 0.97 | 0.97 | 0.97 | 0.95 | 0.99 | 0.97 | 0.98 | 0.98 |
| 1 | 0.9 | 0.91 | 0.84 | 0.94 | 0.97 | 0.96 | 0.92 | 0.96 | 0.93 | 0.92 | 0.94 | 0.96 |

## COI Z PC2 classification by r

| months \ years | 2007 | 2008 | 2009 | 2010 | 2011 | 2012 | 2013 | 2014 | 2015 | 2016 | 2017 | all |
|---|---|---|---|---|---|---|---|---|---|---|---|---|
| 12 | 0.28 | 0.17 | 0.14 | 0.02 | 0.04 | 0.01 | 0.12 | 0.04 | 0.54 | 0.43 | 0.32 | 0.26 |
| 11 | 0.25 | 0.07 | 0.1 | 0.17 | 0.01 | 0.17 | 0.04 | 0.08 | 0.16 | 0.26 | 0.29 | 0.13 |
| 10 | 0.16 | 0.02 | 0.09 | 0.31 | 0.07 | 0.09 | 0.14 | 0.23 | 0.25 | 0.22 | 0.13 | 0.15 |
| 9 | 0.26 | 0.07 | 0.13 | 0.04 | 0.29 | 0.29 | 0.25 | 0.17 | 0.25 | 0.22 | 0.32 | 0.18 |
| 8 | 0.13 | 0.07 | 0.08 | 0.12 | 0.07 | 0.14 | 0.16 | 0.2 | 0.24 | 0.24 | 0.08 | 0.01 |
| 7 | 0.09 | 0.02 | 0.01 | 0.05 | 0.08 | 0.29 | 0.11 | 0.03 | 0.1 | 0.16 | 0.1 | 0.09 |
| 6 | 0.01 | 0.01 | 0.06 | 0.07 | 0.11 | 0.09 | 0.25 | 0.02 | 0.17 | 0.17 | 0.21 | 0.1 |
| 5 | 0.19 | 0.04 | 0.11 | 0.19 | 0.13 | 0.1 | 0.07 | 0.16 | 0.12 | 0.19 | 0.03 | 0.09 |
| 4 | 0.16 | 0.08 | 0.05 | 0.14 | 0.15 | 0.2 | 0.05 | 0.12 | 0.07 | 0.21 | 0.2 | 0.12 |
| 3 | 0.09 | 0.31 | 0.05 | 0.09 | 0.22 | 0.16 | 0.27 | 0.01 | 0.2 | 0.15 | 0.37 | 0.17 |
| 2 | 0.22 | 0.3 | 0.29 | 0.09 | 0.19 | 0.2 | 0.18 | 0.24 | 0.08 | 0.09 | 0.17 | 0.19 |
| 1 | 0.38 | 0.31 | 0.36 | 0.16 | 0.02 | 0.08 | 0.15 | 0.1 | 0.3 | 0.35 | 0.29 | 0.22 |

## COI Z PC3 classification by r

| months \ years | 2007 | 2008 | 2009 | 2010 | 2011 | 2012 | 2013 | 2014 | 2015 | 2016 | 2017 | all |
|---|---|---|---|---|---|---|---|---|---|---|---|---|
| 12 | 0.05 | 0.18 | 0.02 | 0.1 | 0.09 | 0.11 | 0.2 | 0.28 | 0.01 | 0.2 | 0.28 | 0.09 |
| 11 | 0.02 | 0.1 | 0.19 | 0.06 | 0.08 | 0.11 | 0.08 | 0.06 | 0.15 | 0.06 | 0.07 | 0.04 |
| 10 | 0.03 | 0.15 | 0.03 | 0.06 | 0.03 | 0.06 | 0 | 0.02 | 0.07 | 0.02 | 0.04 | 0.01 |
| 9 | 0.09 | 0.07 | 0.04 | 0.09 | 0.06 | 0 | 0.03 | 0.12 | 0.01 | 0.09 | 0.03 | 0.07 |
| 8 | 0.17 | 0.08 | 0.11 | 0.09 | 0.06 | 0.02 | 0.08 | 0.06 | 0.06 | 0.02 | 0.08 | 0.15 |
| 7 | 0.01 | 0.08 | 0.04 | 0.07 | 0.15 | 0.05 | 0.13 | 0.04 | 0.05 | 0.1 | 0.09 | 0.09 |
| 6 | 0.02 | 0.09 | 0.04 | 0.13 | 0.06 | 0.07 | 0.06 | 0.08 | 0.03 | 0.04 | 0.08 | 0.04 |
| 5 | 0.05 | 0.05 | 0.05 | 0.14 | 0.03 | 0.06 | 0.13 | 0.03 | 0.03 | 0.04 | 0.08 | 0.07 |
| 4 | 0.06 | 0.16 | 0 | 0.08 | 0.03 | 0.07 | 0.11 | 0.04 | 0.02 | 0.16 | 0.07 | 0.06 |
| 3 | 0.07 | 0.03 | 0.1 | 0.11 | 0.13 | 0.11 | 0.06 | 0.06 | 0.07 | 0.21 | 0.03 | 0.01 |
| 2 | 0.03 | 0.01 | 0.14 | 0.11 | 0.1 | 0.1 | 0.13 | 0.1 | 0.11 | 0.06 | 0.06 | 0.04 |
| 1 | 0.14 | 0.19 | 0.16 | 0.28 | 0.17 | 0.03 | 0.22 | 0.24 | 0.1 | 0.18 | 0.18 | 0.11 |

**Figure S5.2**. Same as Figure S5.1 but for the Z component.

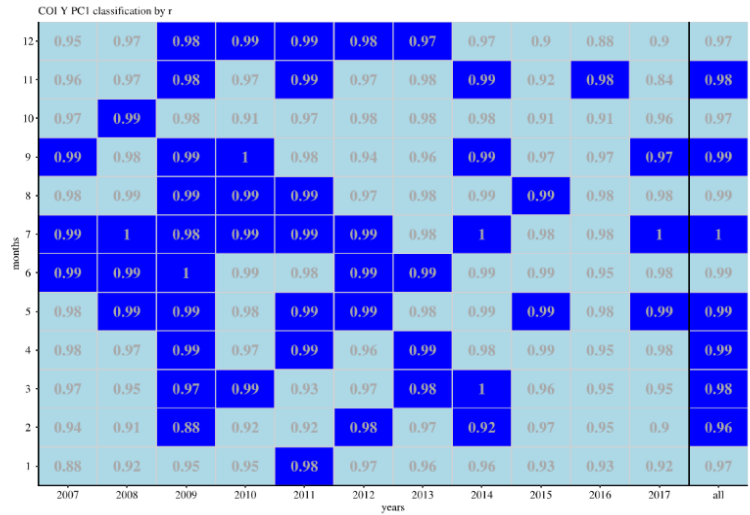
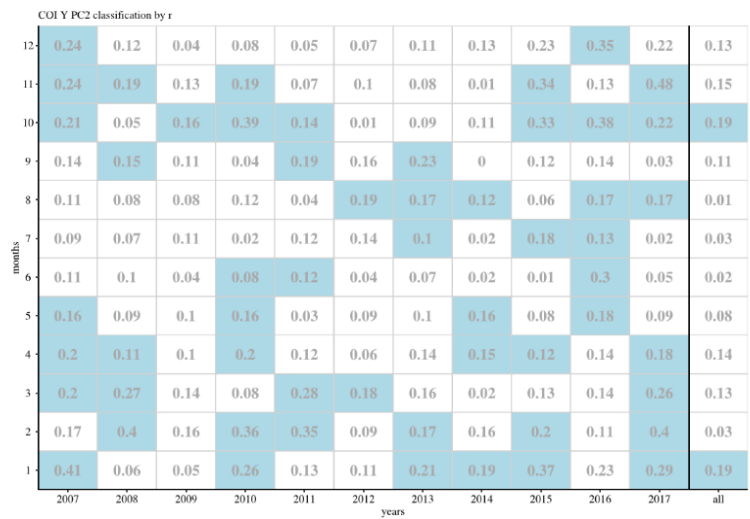
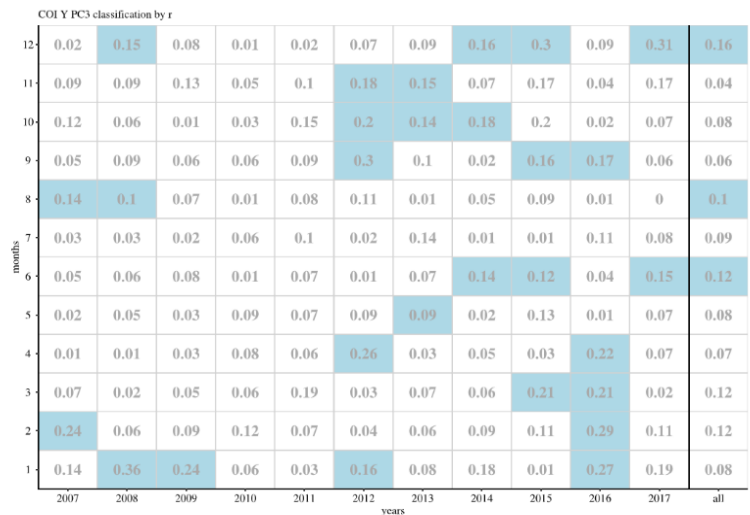

**Figure S5.3.** Combined classification for the Y component: correlation coefficients between the $Sq_{IQD}$ and PC1 (top), PC2 (middle) and PC3 (bottom) series for different months (Y-axis) and different years (X-axis). Blue tiles mark PCs classified as Sq and light blue tiles mark PCs classified as Sq in pair with another PC. (compare with Fig. 2 of the Main Text).

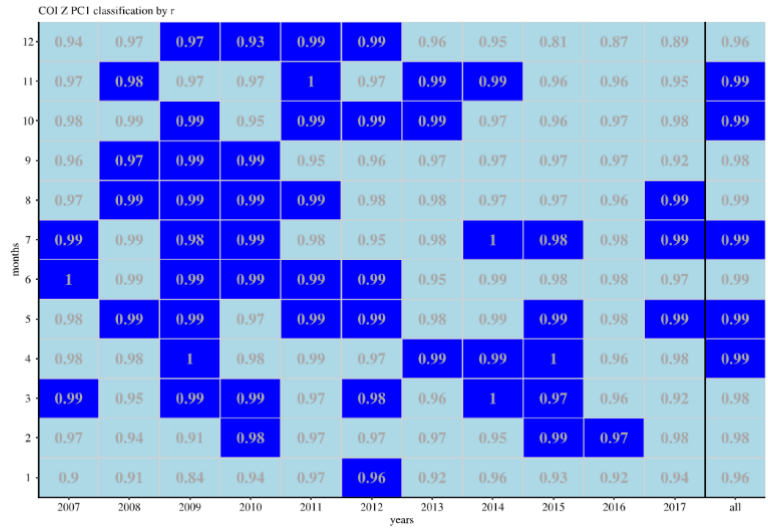
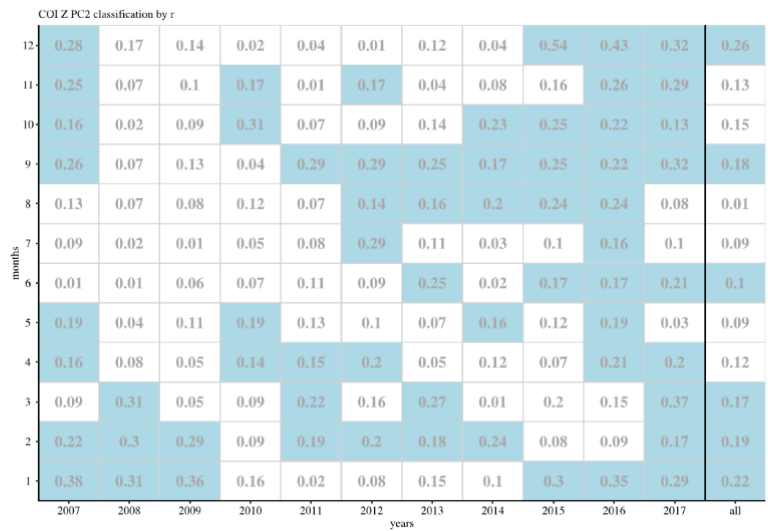
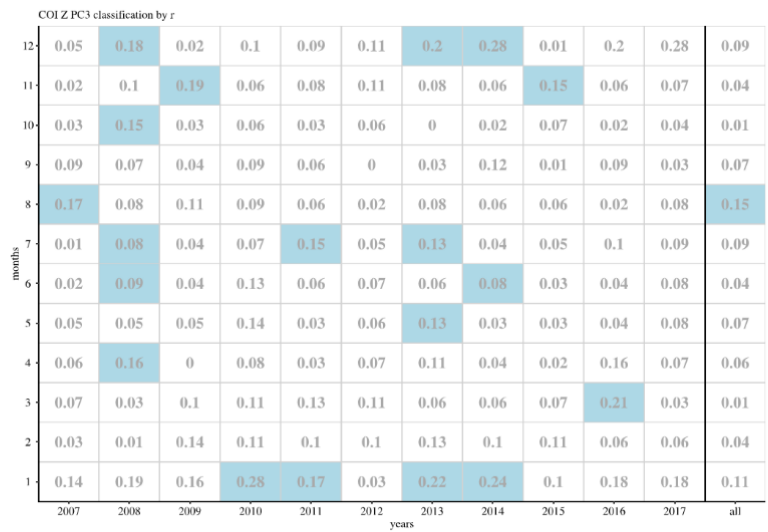

**Figure S5.4.** Same as Figure S5.3 but for the Z component.

Principal component analysis as a tool to extract Sq variation from the geomagnetic field observations: conditions of applicability

Anna Morozova, Rania Rebbah

Supplementary Material, file SM**6**

**Classification of PCs for the X component with Sq$_{CM5}$ as the reference series**

**Figures S6.1-S6.4**

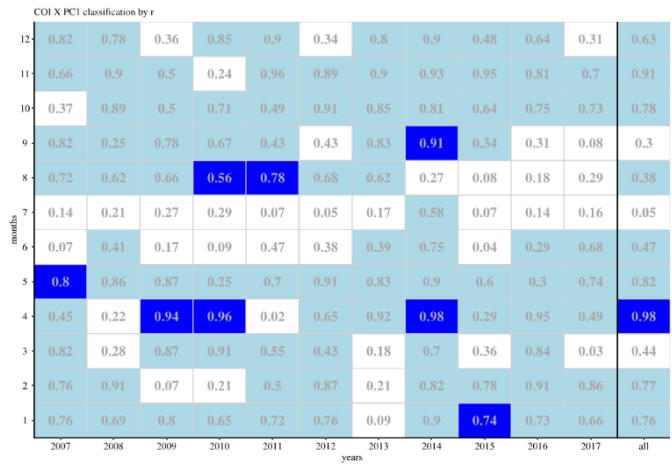
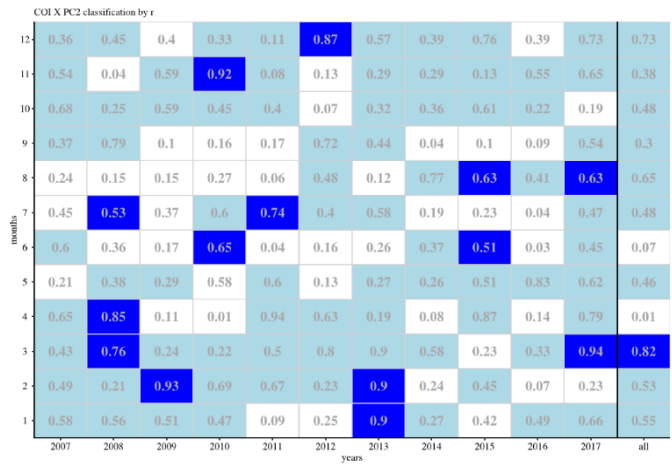
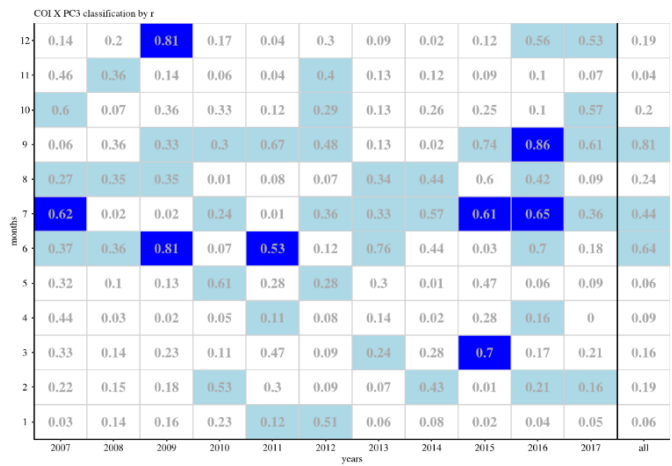

**Figure S6.1**. Combined classification for the X component with Sq$_{CM5}$ as a reference series and *r* is the classification parameter: correlation coefficients (numbers) between the Sq$_{CM5}$ and PC1 (top), PC2 (middle) and PC3 (bottom) series for different months and different years. Blue tiles mark PCs classified as Sq and light blue tiles mark PCs classified as Sq in pairs with another PC (see Figure S6.2). Can be compared with Figure 14 in the Main Text.

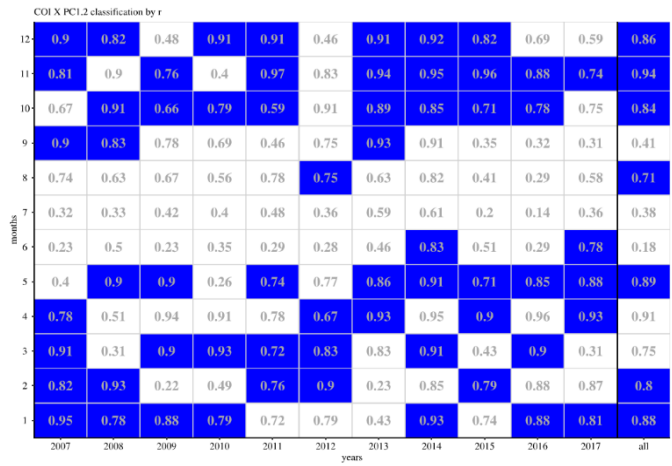
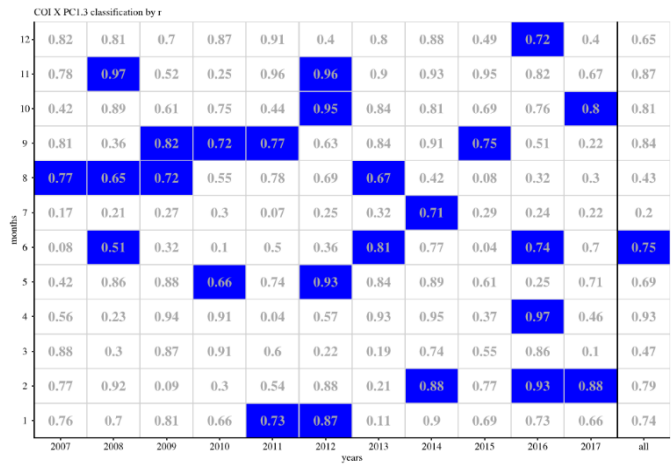
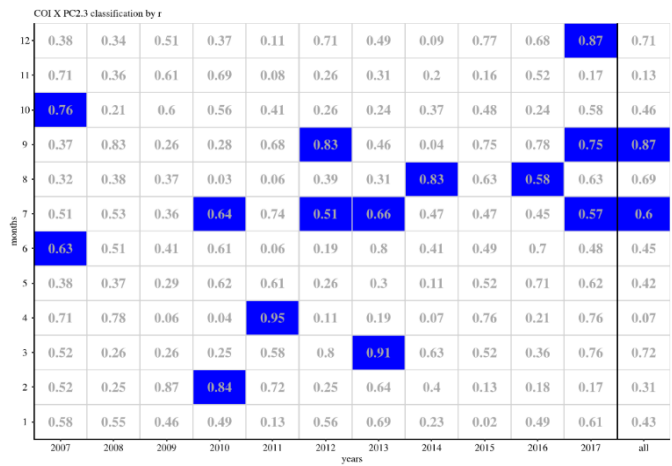

**Figure S6.2**. Same as Fig. S6.1 but for a pair of PCs (top – PC1+PC2, middle – PC1+PC3, bottom – PC1+PC3). Can be compared with Figure 15 in the Main Text.

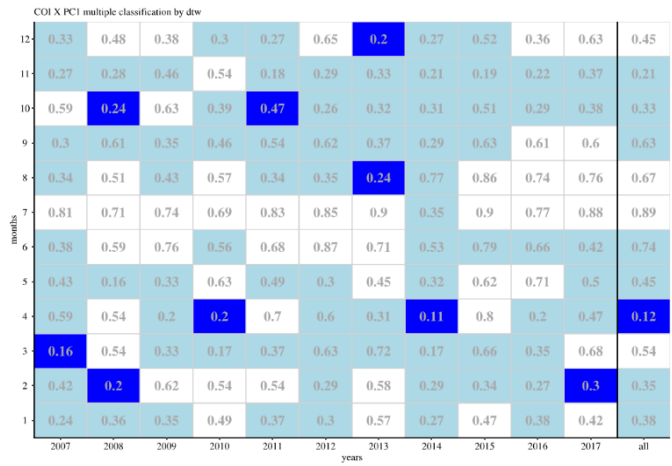
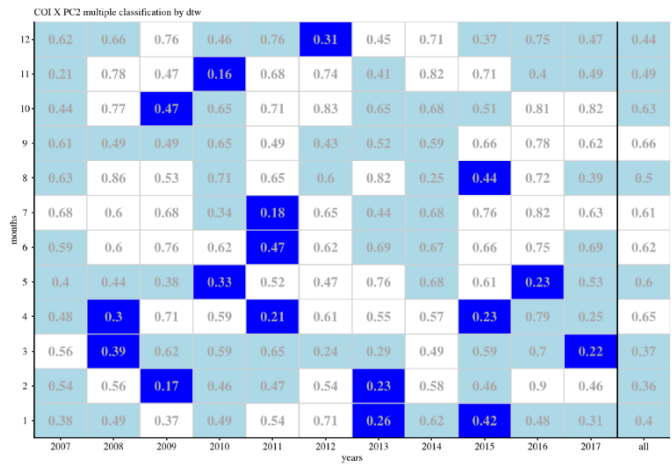
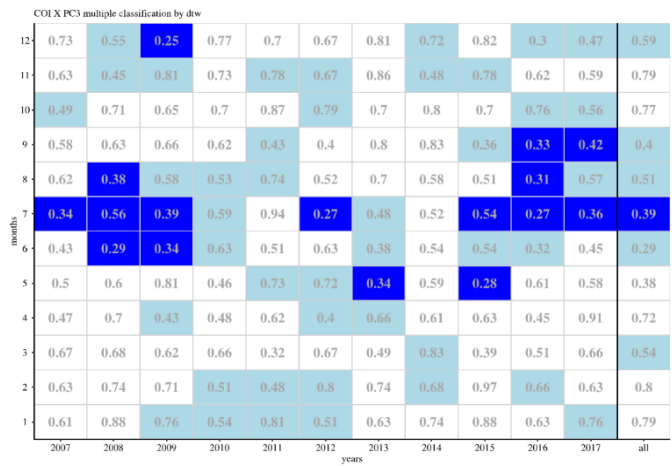

**Figure S6.3**. Combined classification for the X component with Sq$_{CM5}$ as a reference series and *dtw* is the classification parameter: *dtw values* (numbers) between the Sq$_{CM5}$ and PC1 (top), PC2 (middle) and PC3 (bottom) series for different months and different years. Blue tiles mark PCs classified as Sq and light blue tiles mark PCs classified as Sq in pairs with another PC (see Figure S6.4). Can be compared with Figure 16 in the Main Text.

**Figure S6.4.** Same as Fig. S6.3 but for a pair of PCs (top – PC1+PC2, middle – PC1+PC3, bottom – PC1+PC3). Can be compared with Figure 17 in the Main Text.

Principal component analysis as a tool to extract Sq variation from the geomagnetic field observations: conditions of applicability

Anna Morozova, Rania Rebbah

Supplementary Material, file SM7

**Correlation coefficients between the Sq$_{PCA\,(DIFI3,\,dtw)}$ series and the reference series**

**Figure S7.1**

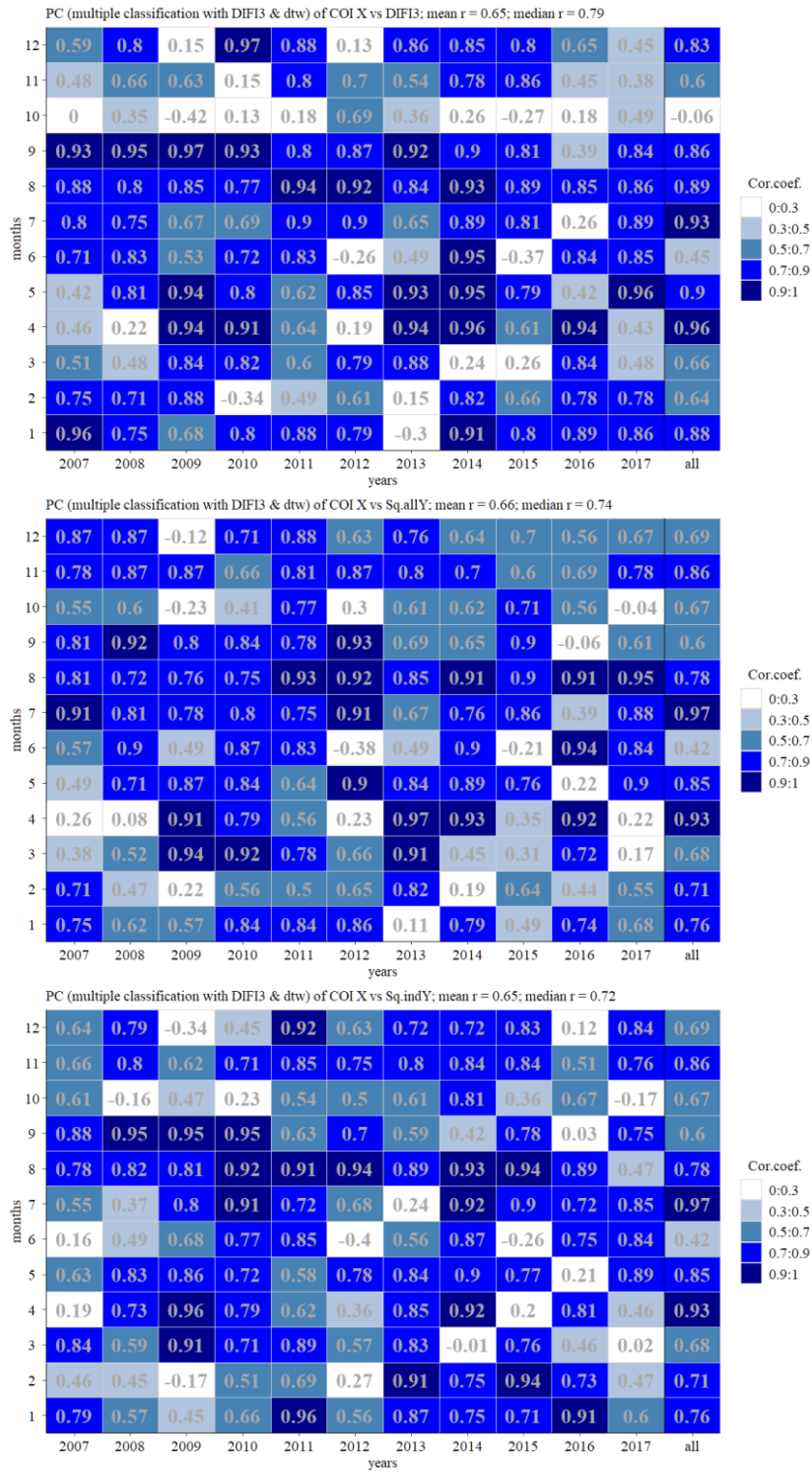

**Figure S7.1**. Correlation coefficients between the Sq$_{PCA}$ series (identified using combined classification with DIFI3 as a reference series and dtw as a metric) and the reference series: Sq$_{DIFI3}$ (top), Sq$_{IQD\ allY}$ (middle) and Sq$_{IQD}$ (bottom).

Principal component analysis as a tool to extract Sq variation from the geomagnetic field observations: conditions of applicability

Anna Morozova, Rania Rebbah

Supplementary Material, file SM**8**

## List of abbreviations

| | |
|---|---|
| Ap, AE, Kp and Dst | indices of geomagnetic activity level |
| COI | Coimbra magnetic Observatory |
| CM5 | the comprehensive model 5 |
| CIRES | Cooperative Institute for Research in Environmental Sciences at the University of Colorado Boulder |
| DIFI3 | The Dedicated Ionospheric Field Inversion model 3 |
| DTW | the dynamic time warping |
| EOFs | empirical orthogonal functions |
| EEJ | the equatorial electrojet |
| ESA | European Space Agency |
| F10.7 | index reflecting variations of the solar UV flux |
| GFZ | German Research Centre for Geosciences at the Helmholtz Centre in Potsdam |
| IQD | international quiet day |
| IAGA | International Association of Geomagnetism and Aeronomy |
| NOC | the natural orthogonal component method |
| PCA | principal component analysis |
| PC(s) | principal component(s) |
| PC1 | 1st principal component |
| PCn | principal component of the n order |
| R | the sunspot number series reflecting variations of the solar UV flux |
| r | correlation coefficient |
| Sq | Solar quiet variation |
| SD | Solar disturbed variation |
| $Sq_{PCA}$ | Solar quiet variations obtained using PCA |
| $Sq_{IQD}$ | Solar quiet variations obtained using IQD |
| $Sq_{CM5}$ | Solar quiet variations obtained using CM5 |
| $Sq_{DIFI3}$ | Solar quiet variations obtained using DIFI3 |
| SVD | The singular value decomposition |
| SCARF | Satellite Constellation Application and Research Facility (SWARM product) |
| VF | variance fraction |

Principal component analysis as a tool to extract Sq variation from the geomagnetic field observations: conditions of applicability

Anna Morozova, Rania Rebbah

Supplementary Material, file SM**9**

## Availability of data and software

The COI 1h data for all geomagnetic components can be downloaded from the World Data Centre for Geomagnetism using the Geomagnetism Data Portal at http://www.wdc.bgs.ac.uk/dataportal/ (station name: "Coimbra", IAGA code: "COI").

The Sq$_{IQD}$ and PCs analyzed in the paper can be downloaded at *Mendeley Data*, doi: 10.17632/jcmdrm5f5x.1, http://dx.doi.org/10.17632/jcmdrm5f5x.1.

All indices of the solar and geomagnetic activity used in this work can be downloaded from the OMNI database at https://omniweb.gsfc.nasa.gov/form/dx1.html.

The CM5 model is available at https://ccmc.gsfc.nasa.gov/models/modelinfo.php?model=CM5.

The DIFI3 model is available at http://geomag.colorado.edu/difi-calculator.

A package for R to calculate *dtw* with different algorithms that were used in this study was developed by Giorgino (2009) and can be downloaded from https://cran.r-project.org/web/packages/dtw/index.html.

.